\documentclass[twocolumn]{aastex62}

\usepackage{graphicx}	          
\usepackage{amsmath}	          
\usepackage{upgreek}            
\usepackage{amssymb}	          
\usepackage{bm}		              
\usepackage{CJKutf8}            
\usepackage{comment}            
\usepackage{tabularx}           
\usepackage{threeparttable}     
\usepackage{booktabs}           
\usepackage{soul}               

\newcommand{\xdex}[1]{\times10^{#1}}
\newcommand{\dex}[1]{10^{#1}}
\def\num#1{\numx#1}\def\numx#1e#2{{#1}\mathrm{e}{#2}}
\newcommand{\fracpartial}[2]{\frac{\partial{#1}}{\partial{#2}}}
\newcommand{\mm}[1]{\mathrm{#1}}
\newcommand{\fracd}[2]{\frac{\mm{d}{#1}}{\mm{d}{#2}}}

\newcolumntype{L}[1]{>{\hsize=#1\hsize\raggedright\arraybackslash}X} 
\newcolumntype{R}[1]{>{\hsize=#1\hsize\raggedleft\arraybackslash}X} 
\newcolumntype{C}[1]{>{\hsize=#1\hsize\centering\arraybackslash}X} 

\received{March 9, 2018}
\revised{May 25, 2018}
\accepted{--}
\submitjournal{ApJ}

\shorttitle{Numerical Robustness of the Streaming Instability}
\shortauthors{Li et al.}

\begin{document}

\title{{\large On the Numerical Robustness of the Streaming Instability:} \\ Particle Concentration and Gas Dynamics in Protoplanetary Disks}

\correspondingauthor{Rixin Li}
\email{rixin@email.arizona.edu}

\author[0000-0001-9222-4367]{Rixin Li
\begin{CJK*}{UTF8}{gbsn}
  (李日新)
\end{CJK*}}
\affil{Steward Observatory \& Department of Astronomy, University of Arizona, \\
 933 N Cherry Ave, Tucson, AZ 85721, USA}

\author{Andrew N. Youdin}
\affiliation{Steward Observatory \& Department of Astronomy, University of Arizona, \\
933 N Cherry Ave, Tucson, AZ 85721, USA}

\author[0000-0002-3771-8054]{Jacob B. Simon}
\affiliation{JILA, University of Colorado and NIST, 440 UCB, Boulder, CO 80309-0440}
\affiliation{Department of Space Studies, Southwest Research Institute, Boulder, CO 80302}
\affiliation{Kavli Institute for Theoretical Physics, UC Santa Barbara, Santa Barbara, CA 93106}



\begin{abstract}
The Streaming Instability (SI) is a mechanism to concentrate solids in protoplanetary disks.  Nonlinear particle clumping from the SI can trigger gravitational collapse into planetesimals.
To better understand the numerical robustness of the SI, we perform a suite of vertically-stratified 3D simulations with fixed physical parameters known to produce strong clumping.
We vary the numerical implementation, namely the computational domain size and the vertical boundary conditions (vBCs), comparing newly-implemented outflow vBCs to the previously-used periodic and reflecting vBCs.
We find strong particle clumping by the SI is mostly independent of the vBCs. However, peak particle densities are higher in larger simulation domains due to a larger particle mass reservoir.  
We report SI-triggered zonal flows, i.e., azimuthally-banded radial variations of gas pressure.  These structures have low amplitudes, insufficient to halt particle radial drift, confirming that particle trapping in gas pressure maxima is not the mechanism of the SI.  
We find that outflow vBCs produce artificially large gas outflow rates at vertical boundaries.  However, the outflow vBCs reduce artificial reflections at vertical boundaries, allowing more particle sedimentation, and showing less temporal variation and better convergence with box size.
The radial spacing of dense particle filaments is $\sim0.15$ gas scale heights ($H$) for all vBCs, which sets the feeding zone for planetesimal growth in self-gravitating simulations. 
Our results validate the use of the outflow vBCs in SI simulations, even with vertical boundaries close  ($\leq 0.4H$) to the disk midplane. 
Overall, our study demonstrates the numerical robustness of nonlinear particle clumping by the SI.
\end{abstract}


\section{Introduction}
\label{sec:intro}

The formation of planetesimals, gravitationally bound solids with sizes $\gtrsim 1$ -- 100 km, is a crucial step in the origin of planetary systems,  leading to the growth of terrestrial planets, giant planet cores, debris disks and Solar System minor planet populations \citep{Youdin2013}.  Many complex physical processes are involved in the growth of planetesimals \citep{Chiang2010, Birnstiel2016}.

Small grains start to grow by collisional coagulation \citep{Blum2008}.  Laboratory experiments combined with theoretical modeling suggest that bouncing and fragmentation pose a barrier to collisional growth beyond $\sim$cm-sizes   \citep{Zsom2010}.  The details of collisional growth barriers, and cases where they might be avoided, are being actively studied \citep{Blum2018}.  
	

To get past the halting or slowing of collisional growth, small solids could gravitationally collapse into planetesimals \citep{Safronov1969, Goldreich1973, Youdin2002}.  Gravitational collapse allows solids to bypass  the ``meter-size barrier", named because the radial drift speeds of meter-sized boulders are quite short, $\sim 10^2$  orbits, in a smooth gas disk \citep{Adachi1976}.   More generally, radial drift is fastest when $\tau_{\rm s} = 1$, where $\tau_{\rm s}$, the ratio of a particle's aerodynamic and orbital timescales, increases with a particle's size \citep{Youdin2010}.  The ability of direct gravitational collapse to overcome the drift barrier is opposed by even weak turbulence in the gas disk \citep{Weidenschilling1980, Youdin2011, Takahashi2016}.

The streaming instability (SI) is an aerodynamic mechanism to concentrate solids to high densities, which can then facilitate gravitational collapse into planetesimals \citep{Youdin2005}.  The SI arises spontaneously from radial drift and two-way  drag forces between solids and gas in protoplanetary disks \citep{Youdin2005, Squire2018}.  
While initial analytic and numerical SI studies neglected vertical gravity \citep{YJ2007},  most current SI simulations (including those presented here) self-consistently include vertical-stratification and shear.  

Strong particle concentration by the SI occurs if many particles have $\tau_{\rm s}$ values are near unity \citep{JY2007, Bai2010a}, which in turn requires significant collisional growth.  More modest particle concentration is possible for $\tau_{\rm s} = 10^{-3}$ in a disk with a large abundance of solids \citep{Yang2017}.  The relevant solid abundance, $Z$, is the locally-averaged ratio of particle-to-gas surface densities.  Strong clumping requires $Z \gtrsim 1\%$ \citep{Johansen2009a}, with the precise value depending on $\tau_{\rm s}$ values \citep{Carrera2015} and increasing with the radial pressure gradient that drives radial drift \citep{Johansen2007a, Bai2010b}.

High-resolution SI simulations that include self-gravity find a broad mass distribution of planetesimals \citep{Simon2016, Schafer2017}, with the surprising finding that variations of $\tau_{\rm s}$ and $Z$ values (within a range that produces strong particle clumping) has a modest effect on this initial mass distribution of planetesimals \citep{Simon2017}.


In addition to understanding how physical parameters affect the SI, it is equally important to understand -- and try to minimize -- the effect of numerical parameters and algorithms on the results of SI simulations.  Particle concentration by the SI on small scales is well-known to increase with the numerical resolution of the hydrodynamic grid \citep{JY2007, Bai2010}, which (in self-gravitating simulations) also extends the low mass end of mass distributions \citep{Johansen2015, Simon2016}.  

With finite computational resources, the resolution of the hydrodynamic grid (as well as the number of particles per grid cell) must compete against the size of the computational domain.  Smaller domain sizes risk introducing artificial interactions, either across the in-plane shear-periodic boundaries of a local disk patch or at vertical boundaries too close to the disk midplane.

To address numerical issues, \citet{Yang2014} used the PENCIL code to simulate  the SI in computational domains of different sizes.  They found that the azimuthally extended particle filaments produced by the SI  have an average radial spacing of 0.2 $H$ (gas scale heights) for their physical parameters.
 
Our study differs from \citet{Yang2014} in two main ways.  First, we use the ATHENA code \citep{Stone2008, Bai2010}.  Second, we also vary the vertical boundary  conditions (vBCs), comparing reflecting, periodic and outflow vBCs.  The outflow vBCs were developed for this study, but also used in \citep{Simon2016, Simon2017}.  We report a similar particle filament spacing (for the same physical parameters) of $\sim 0.15 H$, but using a different, Fourier-space measurement.

Other mechanisms than the SI can concentrate particles, and by enhancing the local particle abundance, $Z$, they may help trigger the SI.   In the ALMA era, the most prominent particle trapping mechanisms are pressure bumps associated with rings or vortices \citep{Pinilla2017}.  A pressure bump created by a planet would occur too late for first-generation planetesimal formation, but would still trap solids \citep{Pinilla2012}.  The magneto-rotational instability is known to produce zonal flows, i.e., radial variations in the orbital speed, which support pressure bumps that can trap particles \citep{Johansen2009, Pinilla2012a}.
	
These large scale pressure bumps  have very different mechanisms than the SI: they are not spontaneously generated by particle-gas interactions.   We report in this paper that the SI also generates small scale zonal flows, which are strongest for outflow vBCs.  However, these SI-induced zonal flows are not strong enough to produce pressure maximum and do not correspond to the location of particle traps.  We discuss the implications of these findings for the interpretation of the SI mechanism.

This paper is organized as follows.  In \autoref{sec:method}, we summarize the numerical model.  We describe  our  implementation of vBCs in \autoref{subsec:vertBCs}  and the parameter space  covered by our simulations in \autoref{subsec:setup}.
We present the results of our simulations \autoref{sec:results}.   Particle sedimentation and peak particle densities are examined in \autoref{subsec:SandC}.  Particle clustering on different length scales is analyzed in \autoref{subsec:Cscale}.  In \autoref{subsec:rings}, we study azimuthally extended  structures, both particle filaments and gas zonal flows.   The vBC-dependent hydrodynamic activity away from the particle-dominated midplane is investigated in \autoref{subsec:flow}.
We summarize and discuss our main conclusions in \autoref{sec:conclusions}.

\section{Method}
\label{sec:method}

To study the coupled dynamics of gas and particles in the midplane of a protoplanetary disk, we use the ATHENA code.  The particle module of ATHENA is described in detail in \citet{Bai2010}.  Subsection \ref{subsec:vertBCs} explains the outflow vBCs that we implemented.  In subsection \ref{subsec:setup}, we summarizes the parameter choices for our simulations. 

\subsection{Equations of Motion}
\label{subsec:EoM}

We simulate a small patch of the protoplanetary disk using the local shearing box approximation, which is useful for studying dynamics on length scales smaller than the disk radius.  With this approximation, the global cylindrical geometry of the disk is mapped onto local Cartesian coordinates with unit vectors $\hat{x}$, $\hat{y}$, and $\hat{z}$ in the radial, azimuthal, and vertical directions, respectively \citep{Goldreich1965}.  The center of the computational domain --- at $(x, y, z) = (0, 0, 0)$ --- is located at a fiducial disk radius, in the midplane.  At this location, the (pressure corrected, see below) Keplerian frequency is $\Omega_0$.  The reference frame orbits, and thus rotates, at this frequency.  The local Keplerian velocity is $v_{\rm K}$.

We solve the equations of continuity and motion for gas, as well as the equations of motion for each solid particle labeled by the subscript $i$:
\begin{align}
\fracpartial{\rho_{\rm g}}{t} + \nabla \cdot (\rho_{\rm g} \bm{u}) &= 0, \label{eq:gascon}\\
\begin{split}\label{eq:gasmom}
\fracpartial{(\rho_{\rm g} \bm{u})}{t} + \nabla\cdot(\rho_{\rm g} \bm{u}\bm{u} + P\bm{I}) &=\\
\rho_{\rm g} \biggl[ 2\bm{u}\times\bm{\Omega}_0 &+ 3{\Omega}_0^2 \bm{x} - {\Omega}_0^2 \bm{z} \biggr] + \rho_{\rm p} \frac{\bar{\bm{v}} - \bm{u}}{t_\mm{stop}},
\end{split} \\
\fracd{\bm{v}_i}{t} = 2\bm{v}_i\times\bm{\Omega}_0+3{\Omega}_0^2 \bm{x}_i &- {\Omega}_0^2 \bm{z}_i-\frac{\bm{v}_i - \bm{u}}{t_\mm{stop}} - 2\eta v_{\rm K} \Omega_0 \hat{x}, \label{eq:ithpar}
\end{align}
where $\rho_{\rm g}$, $\bm{u}$ and $P$ are density, velocity and pressure of gas, $\bm{I}$ is the identity matrix, and $\bm{\Omega}_0 = \Omega_0 \hat{z}$.  For the solids, $\rho_{\rm p}$ and $\bar{\bm{v}}$, give the average density and velocity of the particles in a hydrodynamic grid cell, see \citet{YJ2007} for details.  For individual solid particles, the velocities, $\bm{v}_i$ and radial and vertical positions, $\bm{x}_i$ and $\bm{z}_i$ are indexed.  

The stopping time, $t_{\rm stop}$, measures how quickly drag forces decelerate a particle's motion relative to the gas.  Our models consider a uniform value of the stopping time for all particles.  This choice corresponds to a monodisperse distribution of particle size and mass and to the Stokes or Epstein drag accelerations, which are linear in relative velocity.  

The forces on the gas in \autoref{eq:gasmom} are the Coriolis, radial and vertical tidal gravity (all in square brackets), with the final term giving the back reaction of drag forces on the gas, a crucial ingredient for collective drag effects such as the SI.   The particle equation of motion, \autoref{eq:ithpar}, includes the corresponding acceleration terms for the  solids, with the penultimate term the drag acceleration on a particle.  
The last term in \autoref{eq:ithpar}, $-2\eta v_{\rm K} \Omega_0 \hat{x}$, is a constant inward radial acceleration of particles.  This term replaces the (equal but opposite) outward acceleration of the gas by global radial pressure gradients, which must be included by hand in a local model.  Switching this acceleration to the particles means that $\Omega_0$ is the pressure-supported, slightly sub-Keplerian orbital frequency of the gas.

We adopt an isothermal equation of state, $P = c_{\rm s}^2 \rho_{\rm g}$, where the sound speed, $c_{\rm s}$, is constant.   The units of the code are the natural units of the shearing box.  The time unit is $\Omega_0^{-1}$, the orbital timescale.  The length unit is $H=c_{\rm s}/\Omega_0$, the vertical scale height of the gas.  The mass unit is set by  $\rho_{{\rm g},0}$, the midplane gas density in hydrostatic balance.

\subsection{Vertical Boundary Conditions}
\label{subsec:vertBCs}

\begin{deluxetable*}{cccccccccc}
  \tablecaption{Simulation Parameters\label{tab:paras}}
  \tablecolumns{10}
  \tablewidth{0pt}
  \tablehead{
    \colhead{Run\tablenotemark{*}} &
    \colhead{Domain Size} &
    \multicolumn{5}{c}{Resolution} &
    \colhead{N$_{\rm par}$\tablenotemark{$\dag$}} &
    \colhead{Vertical Flux\tablenotemark{$\ddag$}} &
    \colhead{vBCs} \\
    \colhead{} &
    \colhead{$(L_X\times L_Y\times L_Z)H^3$} &
    \colhead{$N_X$} &
    \colhead{$\times$}  &
    \colhead{$ N_Y$}  &
    \colhead{$\times$}  &
    \colhead{$ N_Z$} &
    \colhead{} &
    \colhead{$\rho_{{\rm g},0}c_{\rm s}$} &
    \colhead{}
  }
  \startdata
  re22  &$0.2\times 0.2\times 0.2$      &$ 64$ &$\times$  &$  64$  &$\times$  &$  64$       &$2^{19}$               &$\cdots       $         &Reflecting    \\
  re24  &$0.2\times 0.2\times 0.4$      &$ 64$ &$\times$  &$  64$  &$\times$  &$ 128$       &$2^{19}$               &$\cdots       $         &Reflecting    \\
  re28  &$0.2\times 0.2\times 0.8$      &$ 64$ &$\times$  &$  64$  &$\times$  &$ 256$       &$2^{19}$               &$\cdots       $         &Reflecting    \\
  re42  &$0.4\times 0.4\times 0.2$      &$128$ &$\times$  &$ 128$  &$\times$  &$  64$       &$2^{21}$               &$\cdots       $         &Reflecting    \\
  re44  &$0.4\times 0.4\times 0.4$      &$128$ &$\times$  &$ 128$  &$\times$  &$ 128$       &$2^{21}$               &$\cdots       $         &Reflecting    \\
  re48  &$0.4\times 0.4\times 0.8$      &$128$ &$\times$  &$ 128$  &$\times$  &$ 256$       &$2^{21}$               &$\cdots       $         &Reflecting    \\
  re82  &$0.8\times 0.8\times 0.2$      &$256$ &$\times$  &$ 256$  &$\times$  &$  64$       &$2^{23}$               &$\cdots       $         &Reflecting    \\
  re84  &$0.8\times 0.8\times 0.4$      &$256$ &$\times$  &$ 256$  &$\times$  &$ 128$       &$2^{23}$               &$\cdots       $         &Reflecting    \\
  re88  &$0.8\times 0.8\times 0.8$      &$256$ &$\times$  &$ 256$  &$\times$  &$ 256$       &$2^{23}$               &$\cdots       $         &Reflecting    \\
  \hline
  pe22  &$0.2\times 0.2\times 0.2$      &$ 64$ &$\times$  &$  64$  &$\times$  &$  64$       &$2^{19}$               &$\num{6.42e-3}$         &Periodic      \\
  pe24  &$0.2\times 0.2\times 0.4$      &$ 64$ &$\times$  &$  64$  &$\times$  &$ 128$       &$2^{19}$               &$\num{4.54e-3}$         &Periodic      \\
  pe28  &$0.2\times 0.2\times 0.8$      &$ 64$ &$\times$  &$  64$  &$\times$  &$ 256$       &$2^{19}$               &$\num{2.46e-3}$         &Periodic      \\
  pe42  &$0.4\times 0.4\times 0.2$      &$128$ &$\times$  &$ 128$  &$\times$  &$  64$       &$2^{21}$               &$\num{5.52e-3}$         &Periodic      \\
  pe44  &$0.4\times 0.4\times 0.4$      &$128$ &$\times$  &$ 128$  &$\times$  &$ 128$       &$2^{21}$               &$\num{5.03e-3}$         &Periodic      \\
  pe48  &$0.4\times 0.4\times 0.8$      &$128$ &$\times$  &$ 128$  &$\times$  &$ 256$       &$2^{21}$               &$\num{1.88e-3}$         &Periodic      \\
  pe82  &$0.8\times 0.8\times 0.2$      &$256$ &$\times$  &$ 256$  &$\times$  &$  64$       &$2^{23}$               &$\num{3.78e-3}$         &Periodic      \\
  pe84  &$0.8\times 0.8\times 0.4$      &$256$ &$\times$  &$ 256$  &$\times$  &$ 128$       &$2^{23}$               &$\num{5.12e-3}$         &Periodic      \\
  pe88  &$0.8\times 0.8\times 0.8$      &$256$ &$\times$  &$ 256$  &$\times$  &$ 256$       &$2^{23}$               &$\num{1.36e-3}$         &Periodic      \\
  \hline
  ou22  &$0.2\times 0.2\times 0.2$      &$ 64$ &$\times$  &$  64$  &$\times$  &$  64$       &$2^{19}$               &$\num{1.78e-3}$         &Outflow       \\
  ou24  &$0.2\times 0.2\times 0.4$      &$ 64$ &$\times$  &$  64$  &$\times$  &$ 128$       &$2^{19}$               &$\num{2.02e-3}$         &Outflow       \\
  ou28  &$0.2\times 0.2\times 0.8$      &$ 64$ &$\times$  &$  64$  &$\times$  &$ 256$       &$2^{19}$               &$\num{1.70e-3}$         &Outflow       \\
  ou42  &$0.4\times 0.4\times 0.2$      &$128$ &$\times$  &$ 128$  &$\times$  &$  64$       &$2^{21}$               &$\num{1.65e-3}$         &Outflow       \\
  ou44  &$0.4\times 0.4\times 0.4$      &$128$ &$\times$  &$ 128$  &$\times$  &$ 128$       &$2^{21}$               &$\num{2.19e-3}$         &Outflow       \\
  ou48  &$0.4\times 0.4\times 0.8$      &$128$ &$\times$  &$ 128$  &$\times$  &$ 256$       &$2^{21}$               &$\num{1.47e-3}$         &Outflow       \\
  ou82  &$0.8\times 0.8\times 0.2$      &$256$ &$\times$  &$ 256$  &$\times$  &$  64$       &$2^{23}$               &$\num{1.55e-3}$         &Outflow       \\
  ou84  &$0.8\times 0.8\times 0.4$      &$256$ &$\times$  &$ 256$  &$\times$  &$ 128$       &$2^{23}$               &$\num{1.69e-3}$         &Outflow       \\
  ou88  &$0.8\times 0.8\times 0.8$      &$256$ &$\times$  &$ 256$  &$\times$  &$ 256$       &$2^{23}$               &$\num{1.19e-3}$         &Outflow       \\
  \enddata
  \tablecomments{For all runs:  the  solid-to-gas ratio  $Z=0.02$, the  dimensionless particle stopping time $\uptau_{\rm s}=0.314$, the headwind speed is $\eta v_{\rm K} = 0.05 c_{\rm s}$, the (cubic) grid cells are $H/320$ wide, and the run-time is $314\Omega_0^{-1}$ ($=50P$).}
  \tablenotetext{*}{Run names consist of a two letter abbreviation for the vertical BC followed by the first significant digit of the horizontal and then vertical size of the computational box.} 
  \tablenotetext{$\dag$}{The number of particles.  For reference, $2^{19} \approx 5.2\times10^5$, $2^{21} \approx 2.1\times 10^6$ and $2^{23}\approx 8.4\times 10^6$. }
  \tablenotetext{$\ddag$}{The time-averaged gas mass flux through vertical boundaries over the initial strong clumping phase.  More specifically, for each snapshot, the horizontal average of the absolute value of mass flux $\langle | \rho_{\rm g}c_{\rm s} | \rangle$ over both the upper and lower boundaries was obtained.}
\end{deluxetable*}

The radial and azimuthal boundary conditions are the standard shear periodic conditions for the shearing sheet, as implemented for ATHENA in \citet{Stone2010}.   As described in the introduction, one of our goals is to study the effect of different vertical boundary conditions, vBCs.  

The classic or standard BCs for SI simulations in a stratified shearing box are the shearing-periodic boundaries in the radial direction, purely periodic in the azimuthal direction and periodic or reflecting in the vertical direction.  While applying the periodic or reflecting vBCs, simulations are assuming there are an infinite amount of disks stacking up together with even spacing.  Thus, gas wave motion owing to the interplay with the particle layer in the midplane will propagate up and down to affect other disks, which numerically is the disk itself.  This configuration is not the favorable situation.  Physically speaking, all truncated disks are artificial.  However, due to the nature of numerical experiments, it is only computationally feasible to study SI within a local shearing box for now, which means there exist vertical limits based on box size.  Hence, we propose to adopt a new vertical boundary condition to improve gas behaviors.  

Following the Appendix of \mbox{\citet{Simon2011}}, we implement a modified outflow boundary condition to handle gas flow at the boundaries.  It extrapolates gas density outside the upper and lower boundaries via an exponential function into ghost zones.  Take the upper ghost zone as an example, the gas density at $Z_{\rm end}+\delta Z$ is
\begin{equation}
  \rho_{\rm g}(Z_{\rm end}+\delta Z) = \rho_{\rm g}(Z_{\rm end}) \exp{\left(-\frac{(\delta Z)^2}{2H^2}\right)},
\end{equation}
where $Z_{\rm end}$ denotes the $\hat{z}$-coordinates of the physical cells right at the upper box boundary (let us call them parent cells), $\delta Z$ indicates the $\hat{z}$-coordinate excesses of the ghost cells outside the upper boundary.  This extrapolation naturally creates the hydrostatic balance since the vertical gravity takes effect everywhere.  Then for each cell in the ghost zone, its horizontal velocity is directly copied from the corresponding parent cell.  But its vertical velocity, $u_z$, will only be copied when the parent cell has a $u_Z$ that does not induce inflow.  Otherwise, the $u_Z$ of the ghost cell will be set to zero to prevent introducing garbage information.  In such manner, the total amount of the materials in the numerical box will slowly monotone decreasing, but the net outward flow is very tiny in each timestep (see \autoref{tab:paras}).  So we renormalize the total mass in the entire domain after each timestep to make it constant \citep{Ogilvie2012}.  This design robustly maintains the hydrostatic equilibrium of gas (see Appendix \ref{app:hydrostatic}).

\subsection{Numerical Parameters}
\label{subsec:setup}

The physical behavior of our simulations is controlled by three dimensionless parameters, which we hold fixed in order to focus on numerical issues.
First, the dimensionless particle stopping time is $\uptau_{\rm s}\equiv \Omega_0 t_{\rm stop} = 0.314$.   Physically, this choice corresponds to compact solid particles with sizes from 10 cm (at  $\sim 1$ AU) to 1 mm  (at $\sim 100$ AU) \citep{Youdin2010}.
Second, we fix the total solids-to-gas ratio at $Z=0.02$, which sets the strength of drag feedback.  The total gas mass used in $Z$ includes the total vertical column density of gas, $\sqrt{2 \pi} \rho_{{\rm g}, 0} H$, 
which extends beyond the computational domain.  Particles remain close to the midplane and require no such correction. 
Third, the headwind parameter $\eta v_{\rm K} / c_{\rm s} = 0.05$, sets the strength of global radial pressure gradients -- which drive radial drift and streaming --  to a typical value at several AU.

\autoref{tab:paras} summarizes the numerical parameters for our simulations.  For all the runs, the initial vertical density profiles of gas and particles are Gaussians with scale heights of $H$ and $H_{\rm p}=0.025H$, respectively, where $H_{\rm p}$ is defined as the root mean square (RMS) of particle $z_i$ coordinates.  Both gas and particles are initialized with the equilibrium drift velocities in \citet{Nakagawa1986}.  

Following the standard local-shearing-box approach, we apply the periodic BCs in both radial and azimuthal direction with azimuthal shear in the radial boundaries.  For the vertical direction, in addition to previously used periodic and reflecting vBCs, we implement and explore the outflow vBCs.  Under each of these three vBCs, we perform a series of nine runs with different numerical domain sizes, served as a convergence test.  Their horizontal sizes $L_X=L_Y$ span from $0.2H$ to $0.8H$ systematically, while their vertical length $L_Z$ spans the same range in the same way but independently.  Their names in \autoref{tab:paras} are composed of the abbreviation of the vBCs, the first significant digit of the box horizontal length and the first significant digit of the box vertical length.  

In order to minimize the influence of numerical resolution, we keep it fixed at $320$ grid cells per $H$ (cell size $\delta l = 0.003125H$). \citet{Bai2010} claimed that an equal number of particles and grid cells is sufficient for numerical convergence in the development of SI.  Considering the disk stratification, the number of the particles should not depend on the height of the numerical box but the typical initial thickness of particle layer, $2H_{\rm p}$.  Therefore, to reduce the number of free parameters in simulations, the number of particles is set by
\begin{equation}
  N_{\rm par} = n_{\rm p} \cdot N_{\rm cell}^* = n_{\rm p} \times \frac{L_XL_Y\cdot 2H_{\rm p}}{(\Delta x)^3} = 8388608\times \frac{L_XL_Y}{(0.8H)^2}
\end{equation}
where $n_{\rm p}=8$ is the number of particles per cell, $N_{\rm cell}^*$ is the number of effective cells around midplane between $2H_{\rm p}$.  Thus in the biggest run, there are over $8$ million particles in a numerical box consisting of $\sim 16$ million grid cells.  Super-particle approximation is applied in our calculations.  To be more specific with the ATHENA code, we use the semi-implicit particle integrator and a triangular shaped cloud (TSC) scheme to interpolate the particle properties to the grid cell for the need of computing interactions between gas and particles \citep{Bai2010}.  Our simulations are computationally expensive due to high resolution and the huge amount of particles.  Therefore, the ending point of each simulation is set to $314 \Omega_0^{-1}$, which is equivalent to $50P$, where $P$ is the orbital period and $P=2\pi\Omega_0^{-1}$.

\begin{figure*}
  \includegraphics[width=\linewidth]{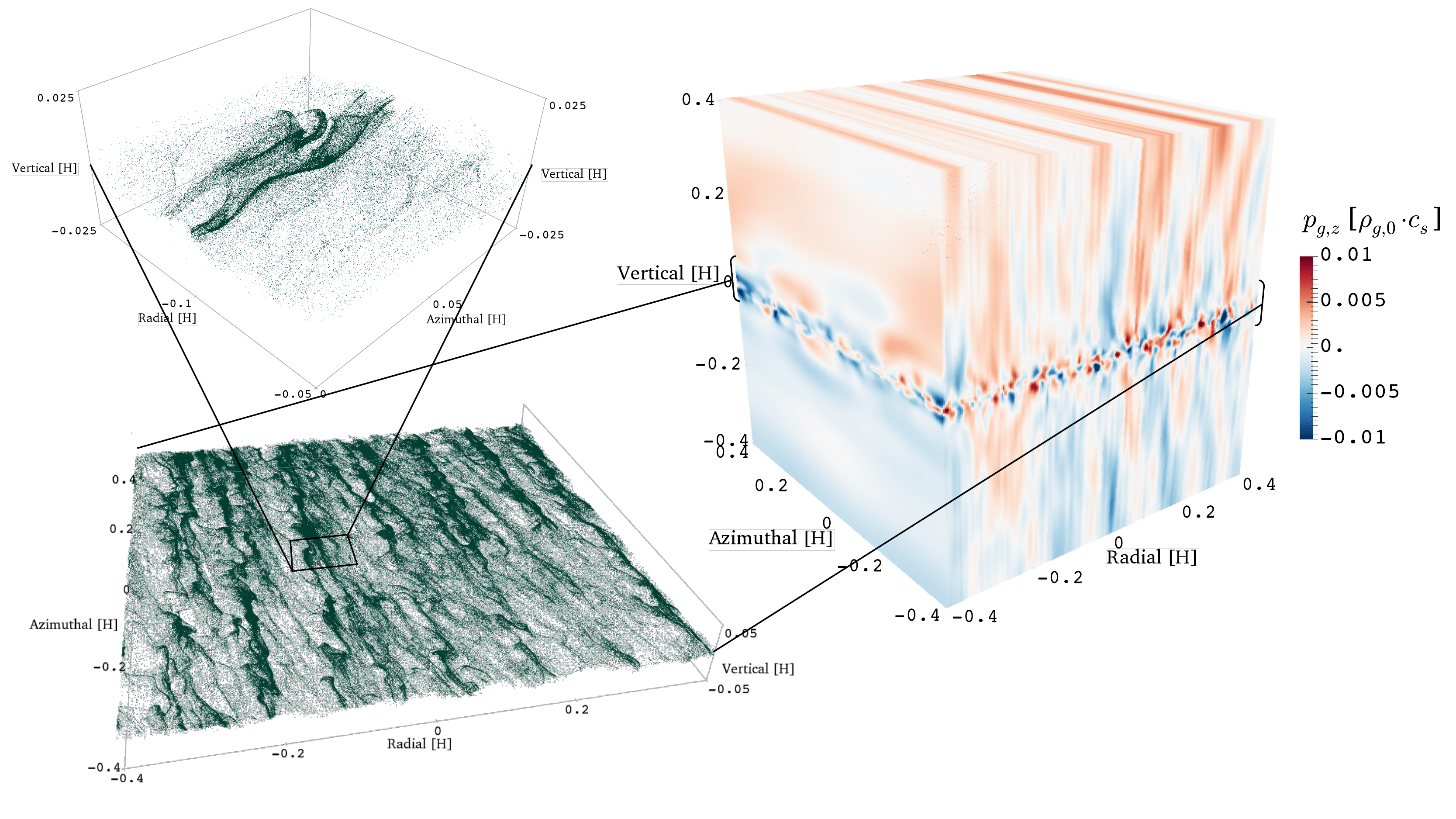}
  \caption{Snapshot of one of our 3D simulations (Run ou88, see \autoref{tab:paras}) at $t = 125\Omega_0^{-1} \approx 20P$.  The color on the box surface (\textit{right}) maps the vertical gas momentum.  Particles in the midplane (\textit{lower left}) form particle filaments due to the streaming instability.  We enlarge a small patch of the simulation domain (\textit{upper left}) to reveal the 3D structures of the particle concentrations.}
  \label{fig:snapshot}
\end{figure*}

\begin{figure*}
  \includegraphics[width=0.495\linewidth]{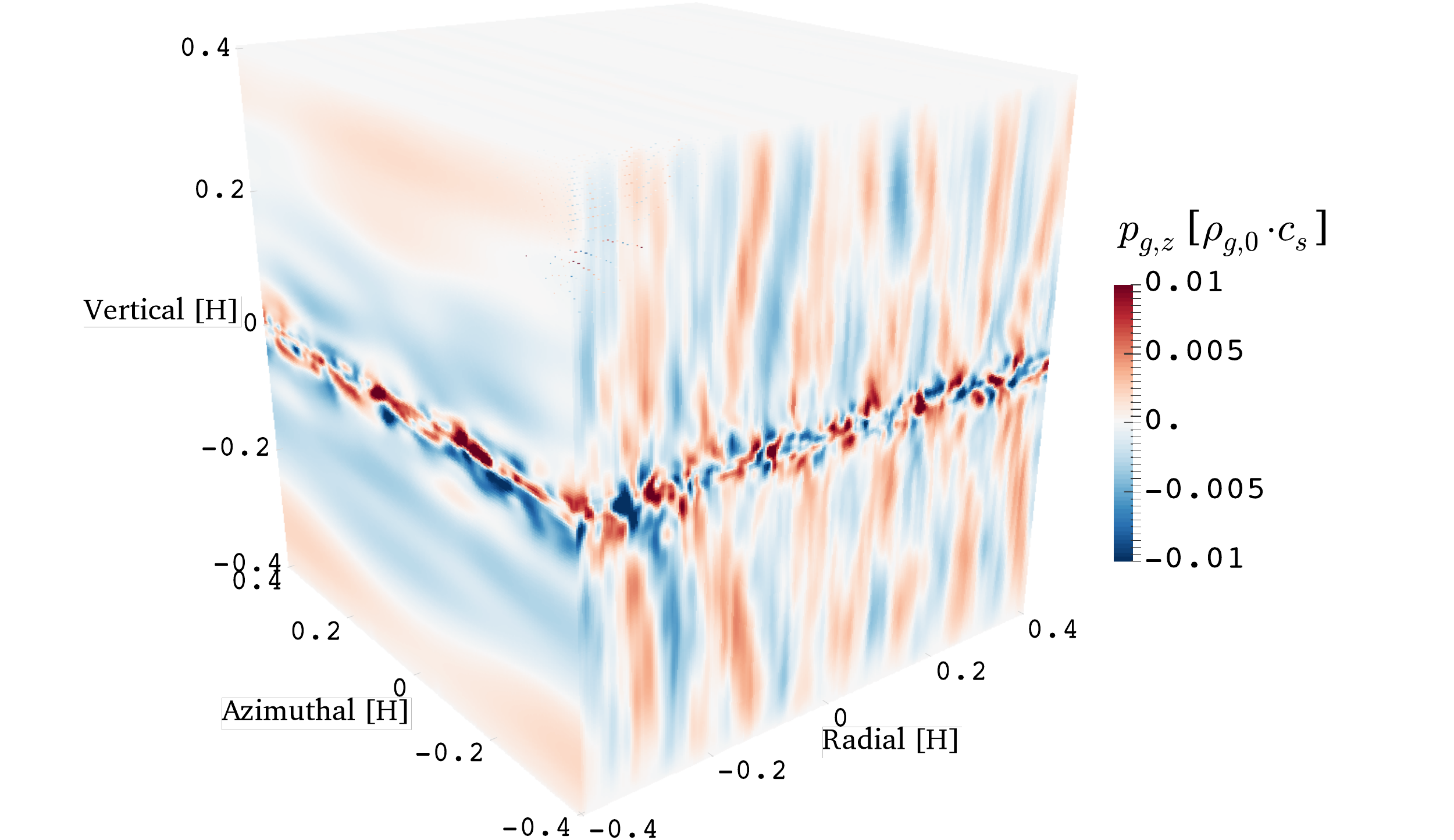}
  \includegraphics[width=0.495\linewidth]{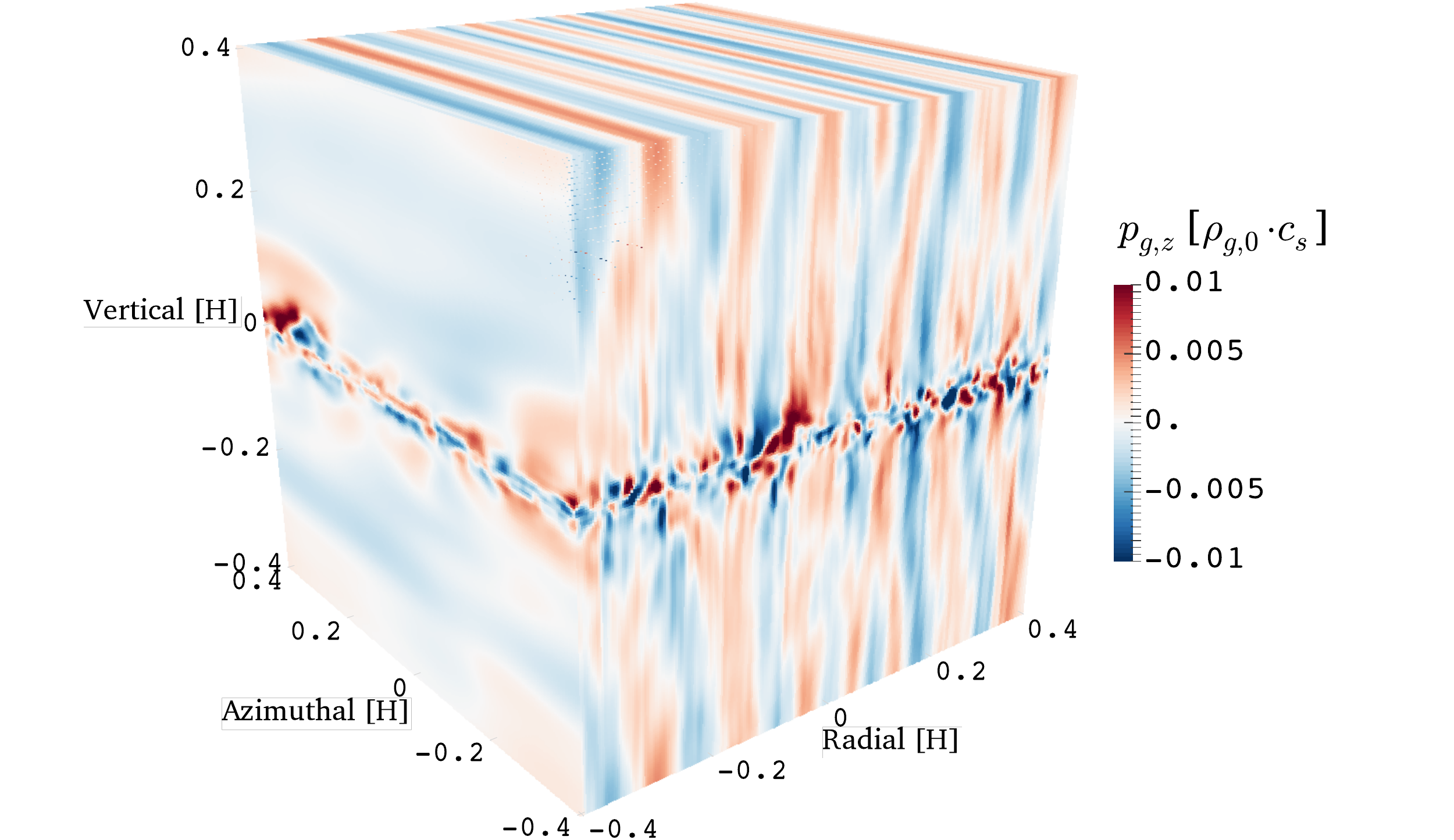}
  \caption{Similar to the right panel of \autoref{fig:snapshot}, but with reflecting vBCs (Run pe88, \textit{left}) and periodic vBCs (Run re88, \textit{right}).}
  \label{fig:perebox}
\end{figure*}


\section{Results}
\label{sec:results}
 
We ran a suite of 3D particle-gas simulations with parameters listed in \autoref{tab:paras}.  The crucial physical parameters were held fixed at values that give strong particle clumping due to SI: $\uptau_{\rm s} = 0.314$, $Z = 0.02$ and $\eta v_{\rm K}/c_{\rm s} = 0.05$.  Our goal is to study the dynamics of this clumping and how it depends on the numerical implementation.  Specifically, we varied the vertical boundary conditions, the horizontal (in-plane) area and the vertical extent of our stratified shearing boxes, negelecting the self-gravity of particles.

\autoref{fig:snapshot} shows a snapshot of Run ou88 at $t=20P$, which uses our new outflow vBCs.  The computational box (\textit{right}) displays slices of the vertical gas momentum. The momentum fluctuations are strongest near the midplane due to particle-gas interactions, but vertically-elongated momentum fluctuations fill the entire box.  The outflow is evident from the positive (negative) vertical momentum on and near the top (bottom) of the box, respectively.  The particle positions (\textit{lower left}) illustrate  strong clumping in a thin midplane layer.  These particle clumps are azimuthally extended with dense knots.  The enlarged panel (\textit{upper left}) zooms in on one of these knots.  

While strong particle clumping exists in all simulations, the response of the gas is different, especially for the different boundary conditions.  For comparison with the vertical outflow case, \autoref{fig:perebox} shows the vertical gas momentum for different vBCs.  The vertically periodic (\textit{left}) case shows alternating inflow and outflow across the vertical boundary.  As required by the periodic BC, outflow from the top (or bottom) becomes inflow into the bottom (or top), respectively.  For the reflecting vBCs (\textit{right}), the vertical flow must vanish at the top and bottom boundaries, but the reflecting case still has similarly strong momentum fluctuations throughout the domain, even a short distance from the vertical boundaries.

We organize our results as follows.  We address the particle sedimentation and clumping in the midplane in \autoref{subsec:SandC} and \autoref{subsec:Cscale}.  We examine the azimuthally extended filaments of both particles and gas in \autoref{subsec:rings}. Finally in \autoref{subsec:flow}, we return to the issue most affected by different vBCs, hydrodynamic flows away from the midplane.

\begin{figure*}
  \centering
  \includegraphics[width= 0.725 \linewidth]{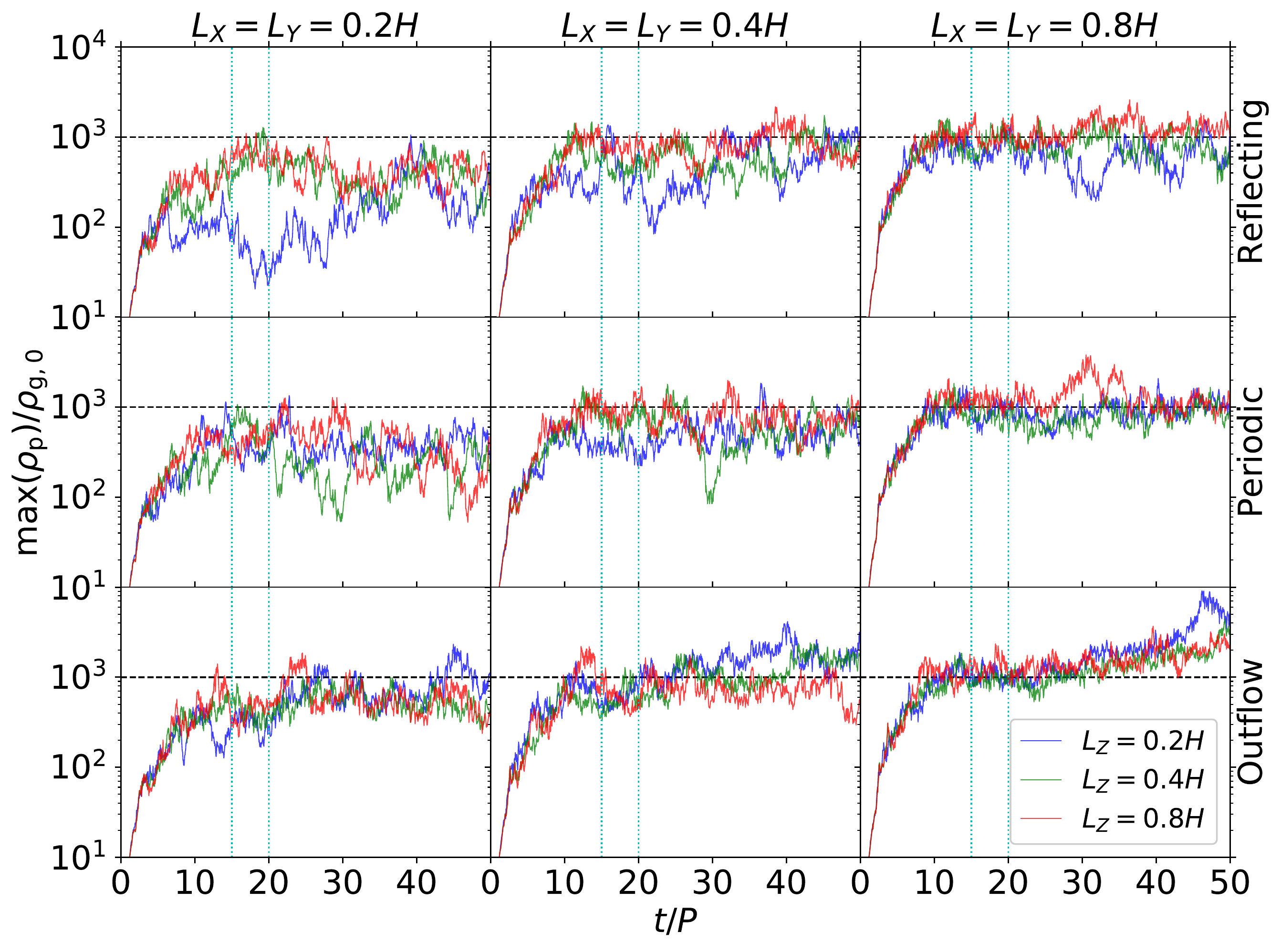}
  \caption{Maximum particle overdensity, max$(\rho_{\rm p})$, as a function of time (in unit of $P=2\pi /\Omega_0$) for all the runs.  These results are divided into three rows by their vBCs.  Columns differentiate horizontal box lengths, $L_X$.  Lines in each frame are color-coded by box heights, $L_Z$.  For better vertical comparison, black dashed reference lines are plotted to represent $\dex{3}\rho_{{\rm g},0}$.  The temporal region sandwiched by two cyan dotted lines are the early strong clumping phases (stage 3) described in \autoref{subsec:SandC}.}
 \label{fig:rhop}
\end{figure*}

\begin{figure*}
  \centering
  \includegraphics[width= 0.725 \linewidth]{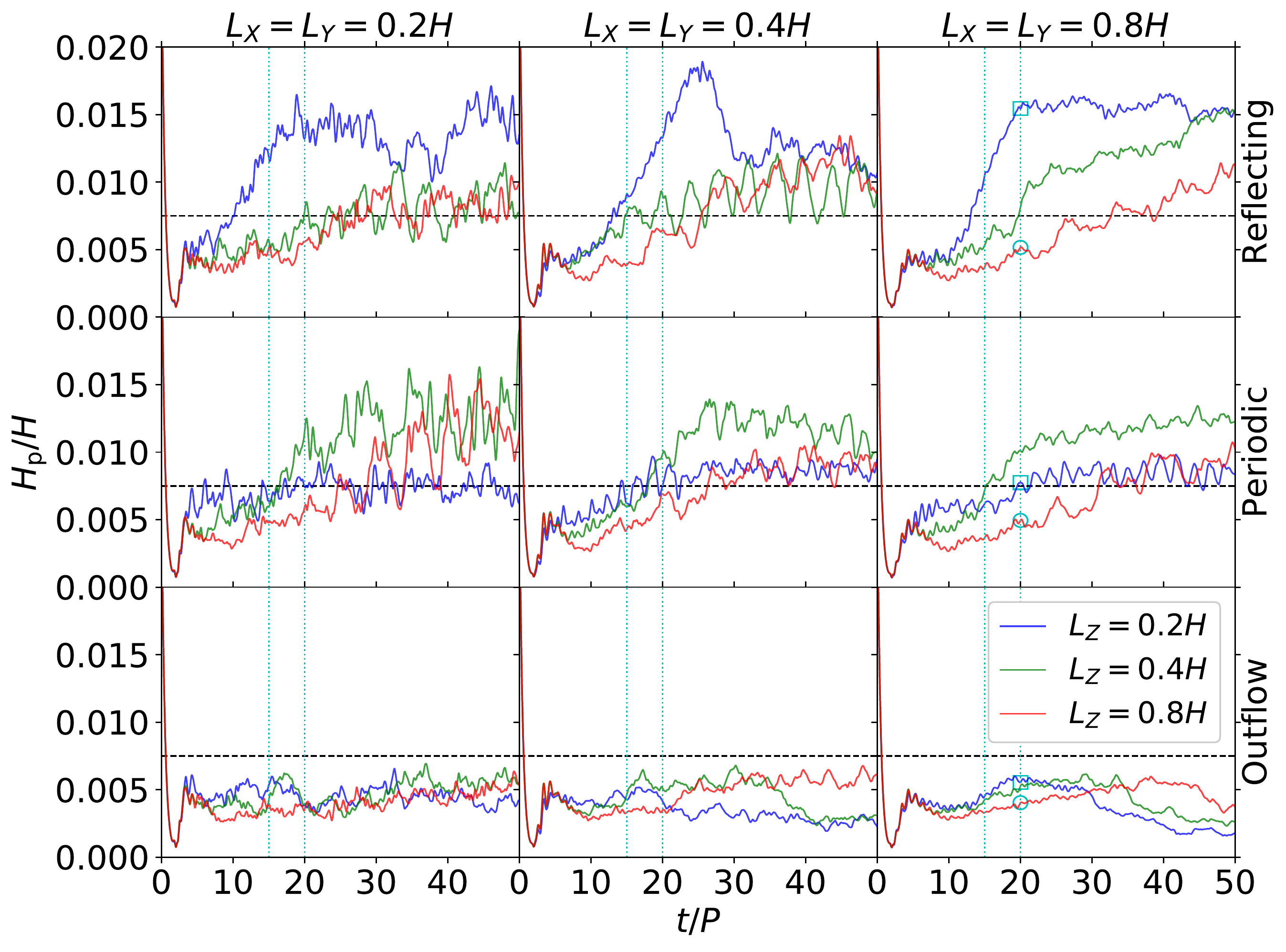}
  \caption{Similar to \autoref{fig:rhop}, but showing particle scale height, $H_{\rm p}$.  For better vertical comparison, black dashed reference lines are plotted to represent $0.0075H$.  Two cyan dotted lines again sandwich the early strong clumping phase.  In the rightmost column, cyan circles mark the simulation states in the largest boxes at $t=20P$, as shown in Figures \ref{fig:snapshot} and \ref{fig:perebox}, and cyan squares mark the model states shown in \autoref{fig:radial_wave}.}
 \label{fig:hp}
\end{figure*}

\begin{figure*}
  \centering
  \includegraphics[width= 0.64 \linewidth]{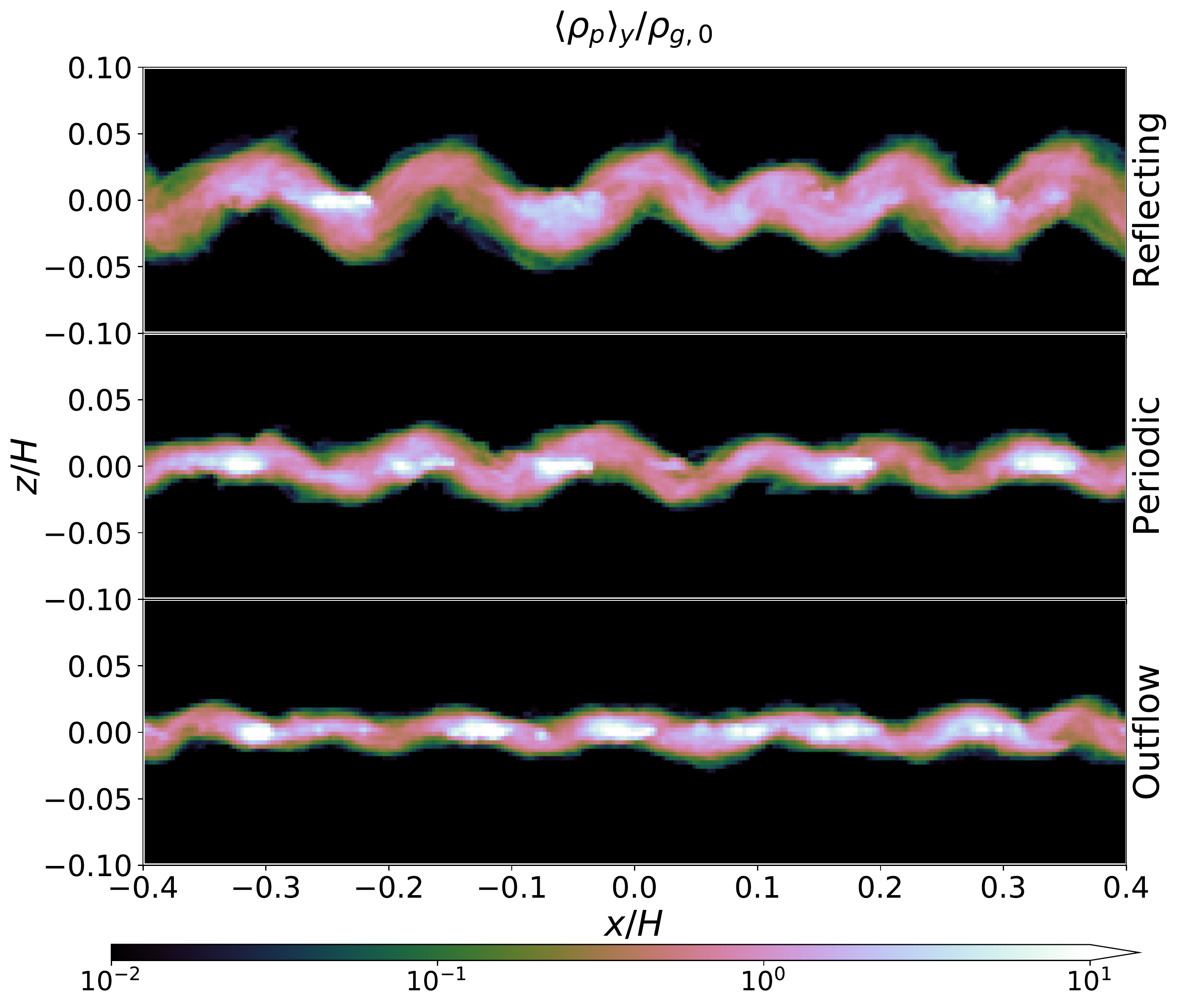}
  \caption{The vertical extent of wave-shaped particle layers. This figure shows the color maps of the azimuthally-averaged particle density, $\langle\rho_{\rm p}\rangle_y$, in radial-vertical ($x-z$) frames, for Run pe82, re82, ou82 (from \textit{top} to \textit{bottom}) at $t=20P$.  In each simulation, SI grows quickly, leading to the formation of azimuthally extended structures in a particle layer.  The value of $H_{\rm p}$ is primarily determined by the thickness of such a particle layer and the amplitudes of the radial waves on it.  Although these three cases have very different $H_{\rm p}$, their max$(\rho_{\rm p})$ are very close to each other at this time (see \autoref{fig:hp} and \autoref{fig:rhop}).  }
 \label{fig:radial_wave}
\end{figure*}

\subsection{Sedimentation and Clustering in the Early Strong Clumping Phase}
\label{subsec:SandC}

The SI is primarily of interest because it can concentrate particles to high densities and facilitate planetesimal formation.  \autoref{fig:rhop} shows the maximum particle density, max$(\rho_{\rm p})$\footnote{As mentioned in \autoref{subsec:setup}, each particle's properties, including density, are smoothly distributed over the nearest three grid cells per dimension by the TSC assignment scheme \citep{YJ2007}.} of all grid cells, as a function of time for all the runs.  

The evolution of max$(\rho_{\rm p})$ is similar for most simulations, and can be divided into four stages.  First, in the sedimentation phase, max$(\rho_{\rm p})$ increases rapidly and steadily to $\sim 20 \rho_{{\rm g},0}$ at $t\sim 2P$.  In the second stage, the streaming instability causes increasingly strong particle clustering, with max$(\rho_{\rm p})$  reaching $\sim \dex{3}\rho_{{\rm g},0}$ in most simulations by $t\sim 15P$. 

We define third stage as lasting from  $t\sim 15P$ to $t\sim 20P$ when peak particle densities  saturate in all our simulations.  The fourth stage, after $t\sim 20P$, has similarly  saturated values of max$(\rho_{\rm p})$, with some fluctuations.  We later show (in \autoref{subsec:rings}) that particle filaments undergo mergers during stage 4 (see also \citealp{Yang2014}).  However, the particle densities in stage 3 are already high enough to trigger gravitational collapse in a self-gravitating simulation.  We thus consider the clustering properties during stage 3 to be the most relevant, especially in terms of initial conditions for gravitational collapse calculations.  We focus on the early strong clumping of stage 3 in most of our analysis.

Our most anomalous simulation is the case of the smallest box ($(0.2H)^3$) with the reflecting vBCs (Run re22, see the blue curve in the upper left frame).  This case shows significantly weaker clumping than all others, especially in stage 3.  Aside from this anomaly, all the simulations exhibit very strong particle clumping.  One apparent trend is that the horizontally largest boxes (with $L_X = L_Y = 0.8H$) show more consistently strong clumping during stage 3.  In the following subsection, we  analyze clumping statistics in more detail.

\autoref{fig:hp} compares the evolution of particle scale heights, $H_{\rm p}$.  In all simulations, $H_{\rm p}$ drops sharply  to $0.001H$ in $t \sim 3P$, as particles sediment. Then,  $H_{\rm p}$ rises sharply  to $\sim0.005H$, the as the SI rapidly develop, with simultaneous particle clustering underway.

After the initial sedimentaion and rebound, the evolution of $H_{\rm p}$ varies significantly with box size and especially vBCs.  With the outflow vBCs, the values of $H_{\rm p}$ remain below $0.0075H$ as indicated by black dashed lines in \autoref{fig:hp}, whereas with the other two vBCs, $H_{\rm p}$ tends to increase and exceed $0.0075H$.  The reflecting vBCs cases evolve to the largest $H_{\rm p}$ values, especially in the shorter $L_Z=0.2H$ boxes.  However, $H_{\rm p}$ is similarly thin during stage 3 in all tall boxes ($L_Z = 0.8H$), largely independent of vBCs or horizontal box size.

To better visualize what causes $H_{\rm p}$ variations in short boxes, \autoref{fig:radial_wave} presents the vertical extent of wave-shaped particle layers.  The corrugated structures of these particle layers are well-illustrated thanks to the axisymmetric particle clumping.  From top to bottom, a larger $H_{\rm p}$ value results from a thicker particle layer as well as the larger amplitudes of the radial waves on it.  

\autoref{fig:radial_wave} also shows that the densest particle clumps (filaments viewed on-axis) lie very close to the midplane. The parts of the corrugated particle layer that are centered away off the midplane have lower particle densities.

\begin{figure*}
  \centering
 \includegraphics[width= 0.99 \linewidth]{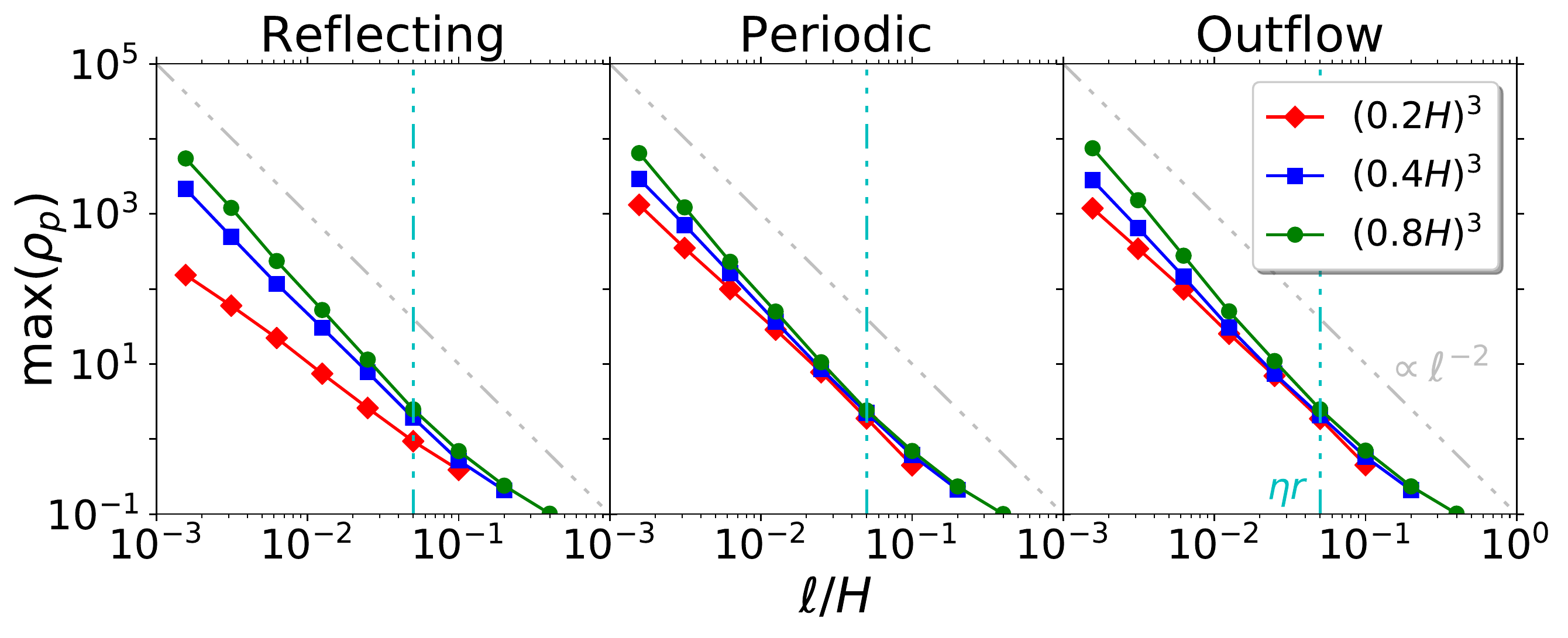}
 \includegraphics[width= 0.99 \linewidth]{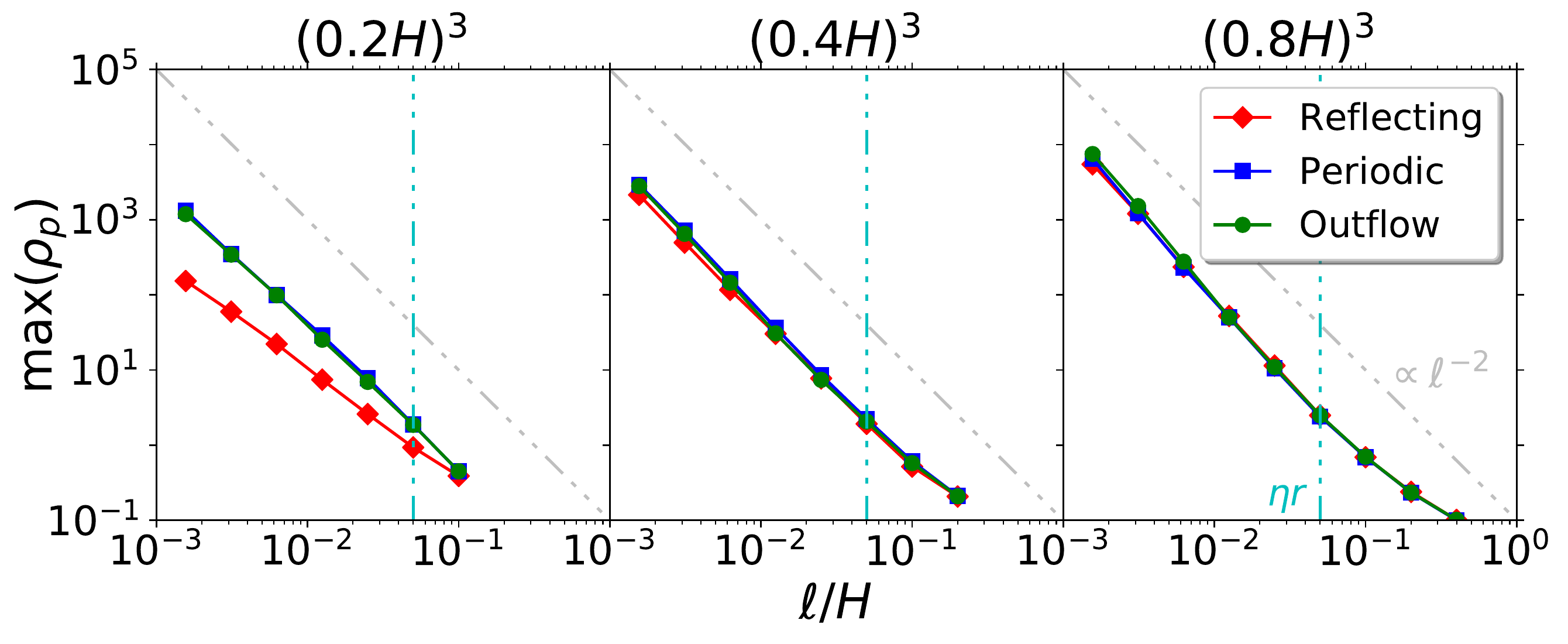}
 \caption{Maximum particle density on a range of lengthscales, time-averaged over the early strong clumping phase, $\langle$max$(\rho_{\rm p})\rangle_t$, for all the runs in cubic boxes.  \textit{Upper}: Each panel shows a different vBC with line colors for different box sizes;  \textit{Lower}: an alternate grouping with each panel showing a given box size and comparing different vBCs.   The diagonal reference lines illustrate a power law model $\langle$max$(\rho_{\rm p})\rangle_t \propto \ell ^{-2}$. The vertical reference lines denote the characteristic lengthscale of \deleted {SI} \added{pressure gradients}, $\eta r$.} 

 \label{fig:rhop_scale}
\end{figure*}

\begin{figure*}
  \centering
 \includegraphics[width= 0.99 \linewidth]{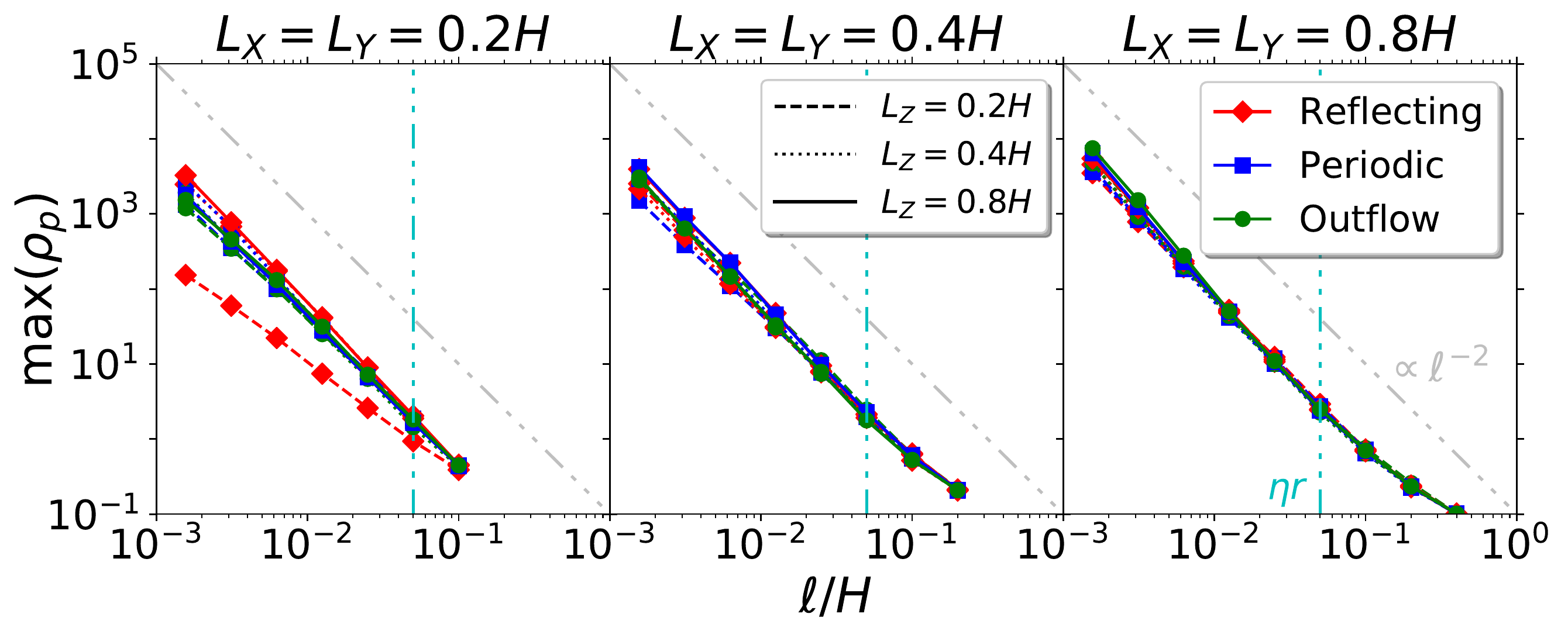}
 \caption{Similar to \autoref{fig:rhop_scale} but with all the cases. These panels are grouped by $L_X(=L_Y)$, with lines colored by vBCs and with line styles according to $L_Z$ values.} 
 \label{fig:rhop_scale_all}
\end{figure*}

\begin{figure*}
  \centering
 \includegraphics[width= 0.99 \linewidth]{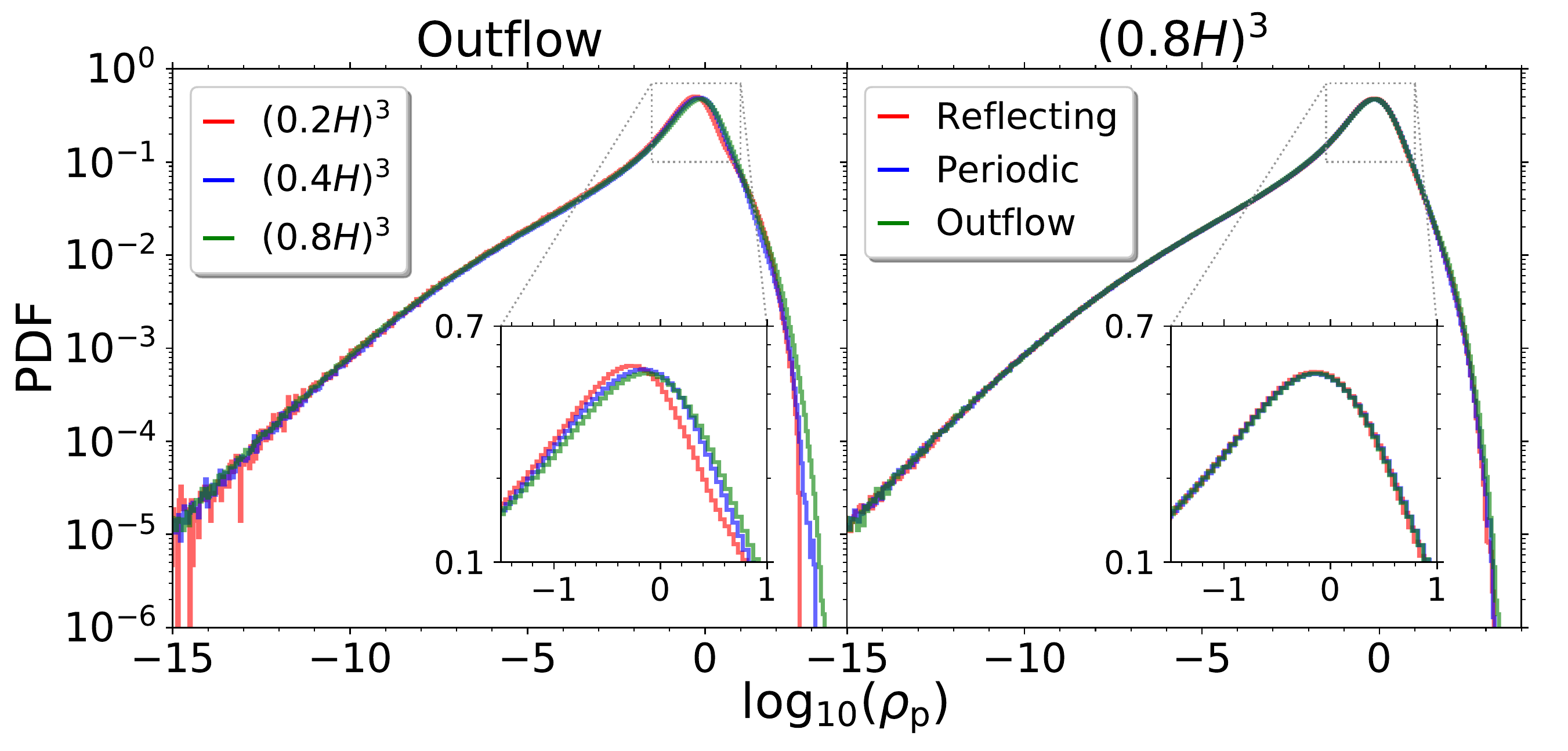}
 \caption{\added{Probability distribution functions (PDFs) of particle density, time-averaged over the early strong clumping phase, with the same choices of runs and colors as the two right panels of \autoref{fig:rhop_scale}.  Regions around the peaks are zoomed in to show the differences. }}
 \label{fig:rhop_PDFs}
\end{figure*}

\subsection{Clustering as a Function of Scale}
\label{subsec:Cscale}

The max$(\rho_{\rm p})$ gives a useful but limited view of the overall particle clustering.  While the peak particle density is useful for estimating whether any fragmentation should occur, it ignores the statistics of particle clustering on different length scales.  Moreover, the results for max$(\rho_{\rm p})$ depend on the cell size and thus do not numerically converge.  For max$(\rho_{\rm p})$, only the densest grid cell is considered, which disregards the rest of the domain and furthermore depends on numerical resolution.   Thus \autoref{fig:rhop_scale} plots the maximum particle density on a range of lengthscales, time-averaged over $t = 15\mbox{--}20 P$.  This scale-dependent $\langle$max$(\rho_{\rm p}) \rangle(\ell)$ gives the highest particle density within a sphere of radius $\ell$ located anywhere in the domain.  Essentially all choices of vBC and box size give the same particle overdensity of $\sim 2 \rho_{\mathrm{g}0}$ at $\ell = \eta r$, the characteristic \added{radial} lengthscale of \replaced{linear SI}{disk pressure gradients and also of the linear (unstratified) SI for $\tau_{\rm s} \simeq 1$ and $\rho_{\rm p} \lesssim \rho_{\rm g}$ (see Fig.\ 2 of \citealp{YJ2007})}.  The only \replaced{exception is the usual anomalous case of}{discrepant case is, again,} the shortest box with reflecting vBCs.  Particle overdensities increase toward smaller scales, down to the resolution limit of the simulations (as shown in \citealp{Johansen2012}).  

The upper panel of \autoref{fig:rhop_scale} compares different box sizes for each of the vBCs.  In all cases, larger boxes produce stronger particle overdensities, especially on small scales.  The larger mass budget in boxes with a larger horizontal area offers the opportunity for larger density fluctuations, and the SI takes advantage.  This horizontal area effect is also seen in \autoref{fig:rhop} and indicates that large boxes are needed to study the maximum extent of small scale clustering.  

The lower panel of \autoref{fig:rhop_scale} presents the effect of vBCs on clumping by simply regrouping the contents in the upper panel.  The clumping of particles across all scales is remarkably independent of the vBCs, again except for the short box reflecting case.  This result is \deleted{very} reassuring for the robustness of particle clustering by the SI since all imposed vBCs are artificial in some way.

\citet{Johansen2012} found that a power law of $\langle$max$(\rho_{\rm p})\rangle_t \propto \ell^{-2}$ (indicated by the diagonal reference line) is a good approximation to the scale-dependence of clumping.  Such a scaling is expected for a linearly elongated structure, whose mass $\propto \ell$ averaged over a volume $\propto \ell^3$ gives density $\propto \ell^{-2}$.  At small scales, this argument breaks down due to the finite width of elongated structures and the presence of overdense knots, as seen in \autoref{fig:snapshot}.  Respectively, these two effects can give either a flatter or a steeper slope for $\langle$max$(\rho_{\rm p})\rangle (\ell)$.  Thus the fact that the slope stays as steep as $-2$, and even gets steeper in the larger boxes, is a sign of strong small-scale clumping\deleted{, well below the lengthscale of the linear SI}.

\added{The existence of strong clumping on length-scales $\ll \eta r$ is consistent with rough expectations of linear SI theory.  As $\mu \equiv \rho_{\rm p}/\rho_{\rm g}$ increases above unity, the radial length scale of the linear SI decreases sharply, while the growth rate increases \citep{Youdin2005, YJ2007, Squire2018}.  These linear properties seem consistent with a cascade of stronger concentration to smaller lengthscales.  Nevertheless, it is difficult to use the linear theory (which is also asixymmetric and unstratified) to predict this non-linear clumping.  Moreover, the box-size dependence of small-scale clustering shows that the shows that small scale clustering is not fully numerically converged,  apparently because it is not a local process.}

\autoref{fig:rhop_scale_all} investigates whether the vertical extent of the computation domain affects particle clustering by comparing runs with the same horizontal area.  Overall, the effect of $L_Z$ on particle clustering is small, especially in the boxes with the largest horizontal area ($L_X = L_Y = 0.8H$).  For all boundary conditions, shorter boxes do have somewhat lower particle overdensities on the smallest scales.  (We again see that the smallest reflecting box (Run re22) is unique among our runs for its weak particle clustering.) Our finding that particle clustering depends only weakly on box height is reassuring since many self-gravitating simulations use short boxes ($L_Z=0.2H$) to reduce computational costs.

\added{\autoref{fig:rhop_PDFs} plots the probability distribution functions (PDFs) of the particle densities at the  grid scale,\footnote{
  \added{The particle densities are interpolated with the TSC scheme, onto cubic cells of width $\delta l$.  The maximum densities in \autoref{fig:rhop_PDFs} are thus comparable, but not exactly equal, to the densities, $\langle$max$(\rho_{\rm p}) \rangle(\ell)$, in a sphere of radius $\ell = 2\ \delta l$ in \autoref{fig:rhop_scale}.}}
again time-averaged over $t = 15\mbox{--}20 P$.  The left panel compares different box sizes for the outflow vBCs.  Larger boxes produce higher peak densities, with PDFs that extend above $10^{3}\rho_{\mathrm{g}0}$.  The right panel shows that vBCs have an impressively negligible effect on density PDFs, in the largest simulation box.  The convergence properties of particle density PDFs on the grid scale thus agrees with the analysis of maximum density  as function of scale.}

\begin{figure*}
  \centering
  \textbf{\Large Outflow vBCs} \\
  \includegraphics[width=0.9 \linewidth]{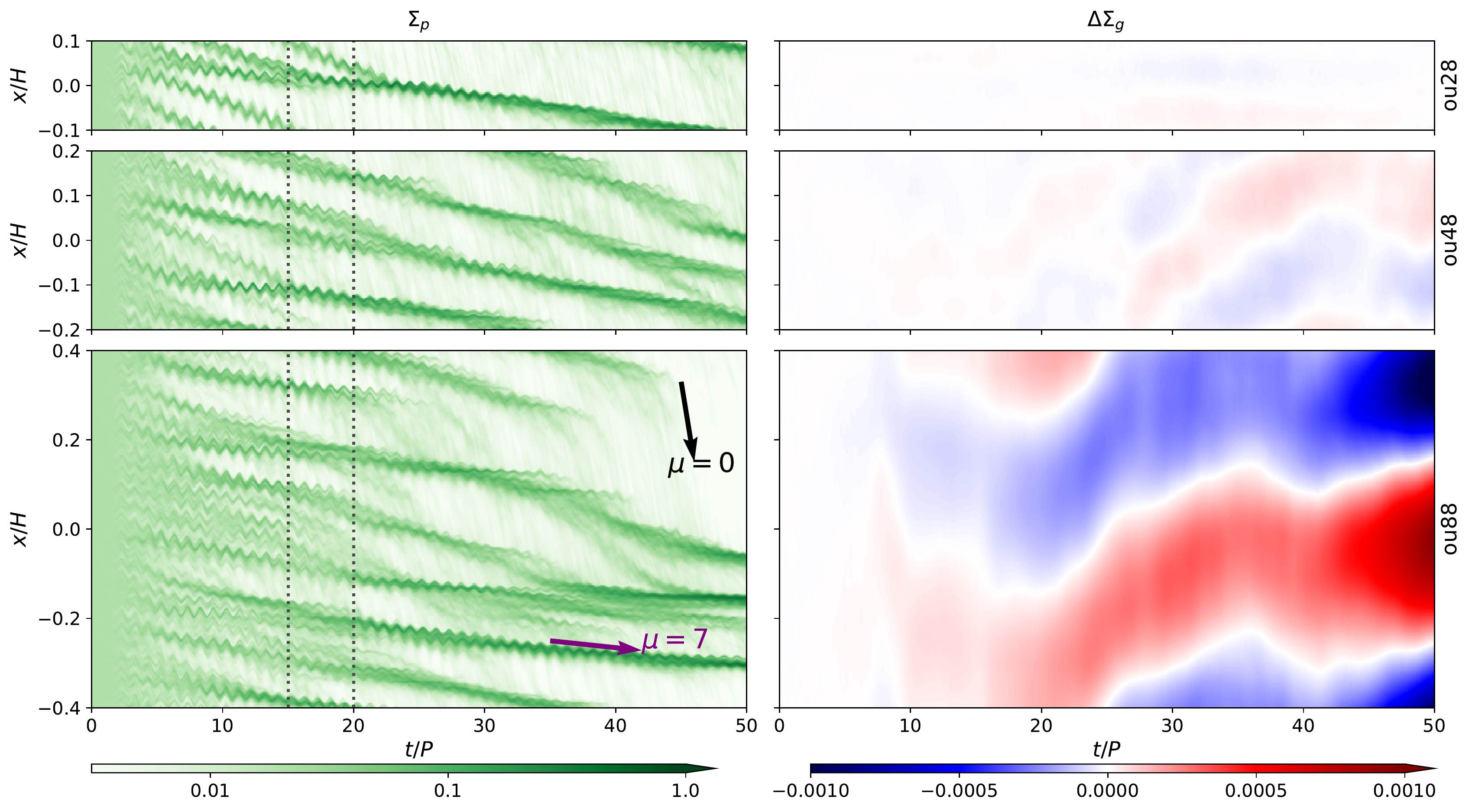}
  \caption{Azimuthally-averaged particle surface density ($\langle\Sigma_{\rm p}\rangle_y/\Sigma_{{\rm g},0}$, \textit{left}) and gas surface density variations ($\Delta\langle\Sigma_{\rm g}\rangle_y/\Sigma_{{\rm g},0}$, \textit{right}) vs. radius ($x$) and time ($t$) for box widths of $L_X=0.2H$ (\textit{top}), $0.4H$ (\textit{middle}), and $0.8H$ (\textit{bottom}),  where $\Sigma_{{\rm g},0}$ is the initial gas surface density.  These runs (ou28, ou48, and ou88) are conducted with the outflow vBCs.  The slope of the black arrow annotated with $\mu = 0$ denotes the drift speed of a test particle, where $\mu \equiv \rho_{\rm p}/\rho_{\rm g}$. The purple arrow shows the equilibrium drift speed for \replaced{$\mu=7.5$}{$\mu = 7$}, which gives an effective value of the particle density in the dense filament that it parallels.  Two vertical dotted lines again sandwich the early strong clumping phase. }
 \label{fig:foSigma}
\end{figure*}

\begin{figure*}
  \centering
  \textbf{\Large Periodic vBCs} \\
  \includegraphics[width=0.9 \linewidth]{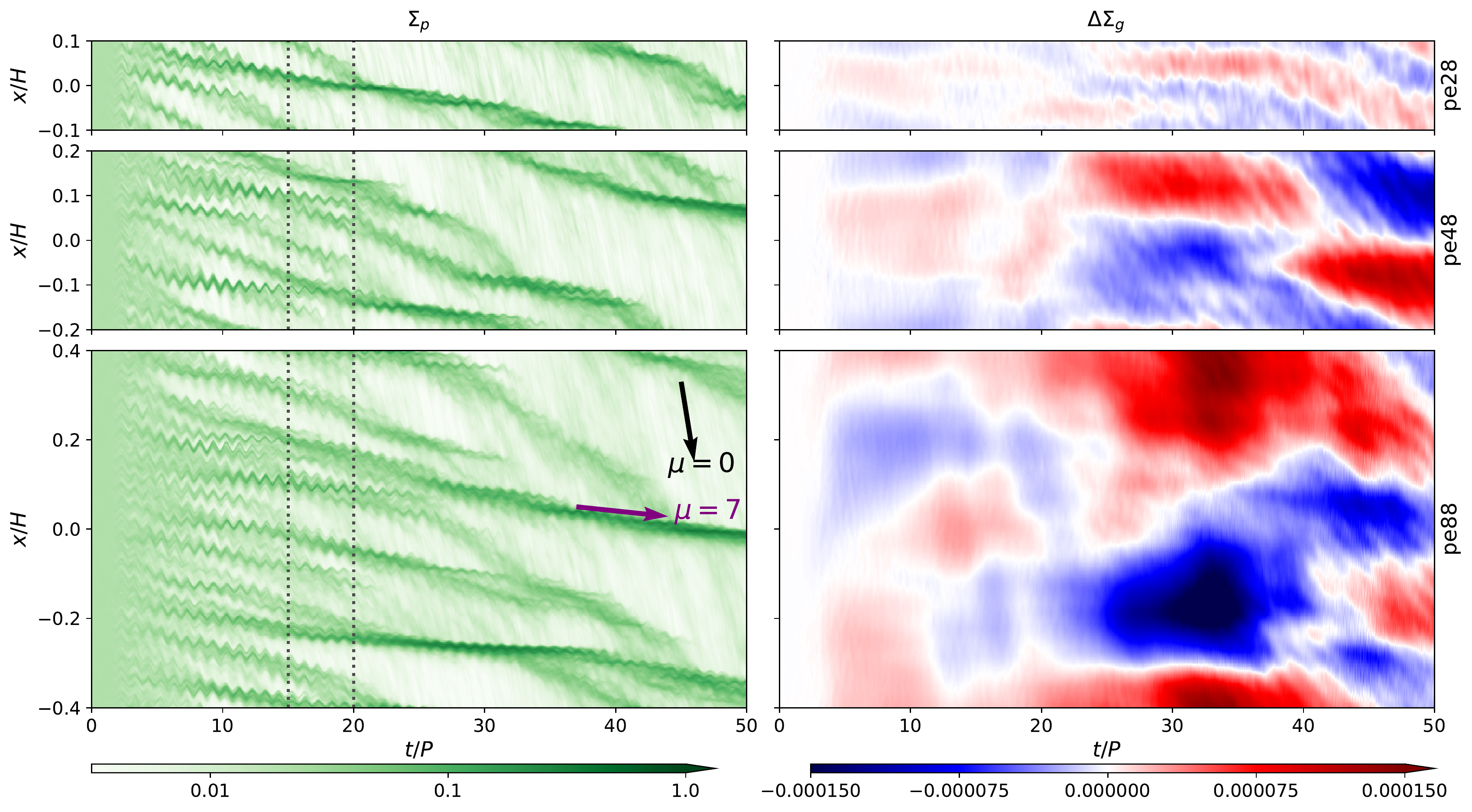}
  \caption{Similar to \autoref{fig:foSigma}, but for Run pe28, pe48 and pe88 that are conducted under periodic vBCs in the vertical direction.  }
 \label{fig:peSigma}
\end{figure*}

\begin{figure*}
  \centering
  \textbf{\Large Reflecting vBCs} \\
  \includegraphics[width=0.9 \linewidth]{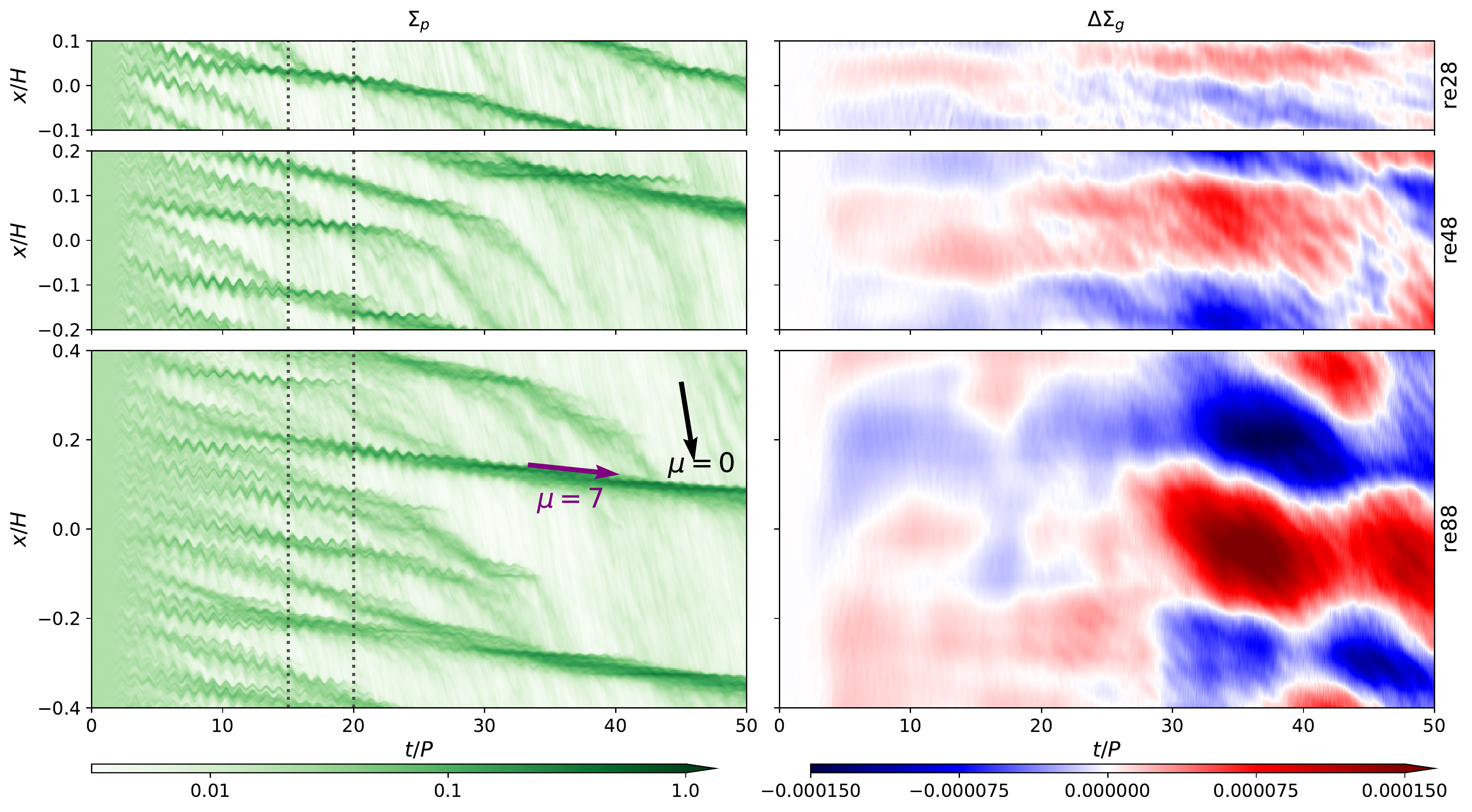}
  \caption{Similar to \autoref{fig:foSigma}, but for Run re28, re48 and re88 that are conducted under reflecting vBCs in the vertical direction.}
 \label{fig:reSigma}
\end{figure*}

\begin{figure*}
  \centering
 \includegraphics[width= 0.8 \linewidth]{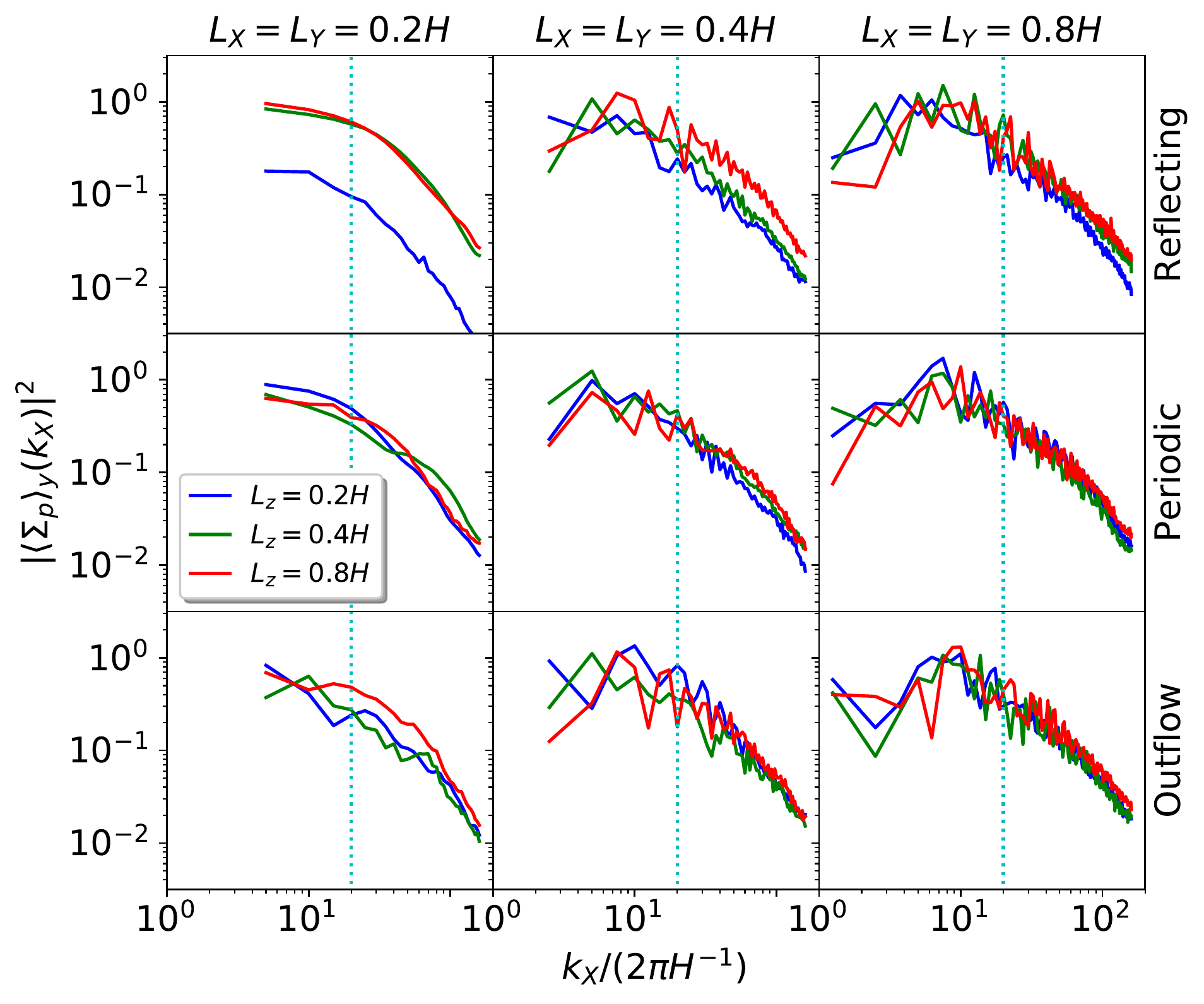}
 \caption{Power spectra of the azimuthally-averaged particle surface density, $|\langle\Sigma_{\rm p}\rangle_y|^2$, for all the runs.  Similar to \autoref{fig:hp}, the results are divided in three rows by the vBCs, grouped in the columns by $L_X$, and color-coded by $L_Z$ (see the legend).  The cyan dotted lines denote the wavenumber corresponding to \deleted{the characteristic length of the SI,} $\eta r$, where $k_X = 2 \pi / (\eta r) = 20\ (2 \pi H^{-1})$. }
 \label{fig:fftpower}
\end{figure*}

\begin{figure}
 \includegraphics[width= \linewidth]{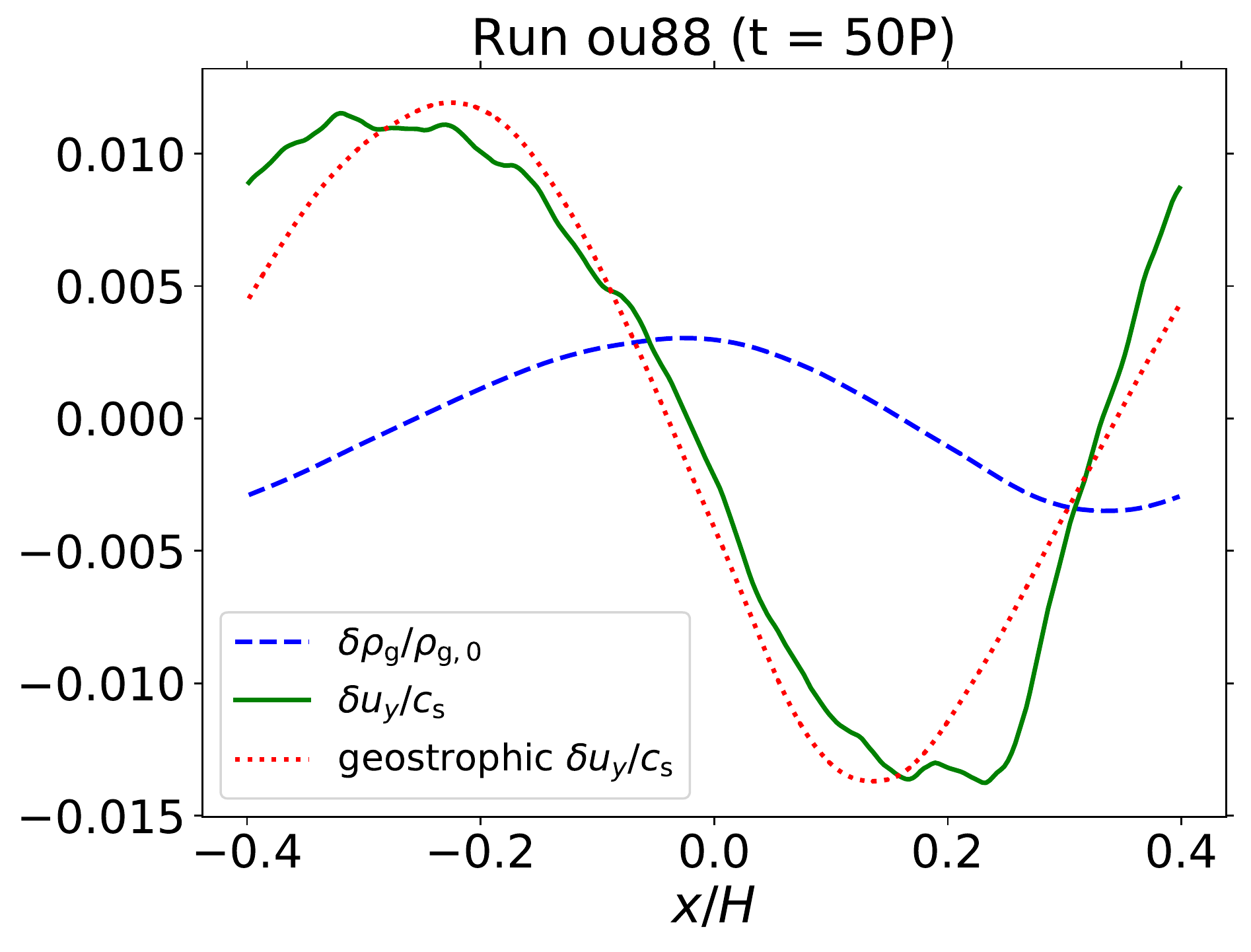}
 \caption{The altazimuthally-averaged perturbations of gas density, $\delta \rho_{\rm g}/\rho_{\mathrm{g},0}$ (\textit{blue dashed line}), and azimuthal gas velocity, $\delta u_y / c_{\rm s}$ (\textit{green solid line}), for Run ou88 at $t=50P$.  The expected $\delta u_y$ from geostrophic balance  (\textit{red dotted line}, from  \autoref{eq:uyi}) matches the amplitude and phase of the zonal flow seen in the simulation.  The amplitude of this zonal flow (the strongest case), $\delta u_y \simeq 0.01c_{\rm s}$ is far to weak to overcome the headwind speed $\eta v_K = 0.05 c_{\rm s}$ and trap particles.}
 \label{fig:zonal}
\end{figure}

\subsection{Azimuthally Extended Particle and Gas Structures}
\label{subsec:rings}

To study the largest structures in our simulations, we analyze the azimuthally-averaged surface density of both particles and gas.  We first focus on the particles to better understand the spacing of azimuthal filaments.  The spacing of these filaments affects the mass reservoir available to form planetesimals in the gravitational fragmentation phase (not modeled here).  We then analyze axisymmetric gas structures to highlight two important points.  First, the SI produces zonal flow structures which, to our knowledge, has not been previously reported, and which depends strongly on horizontal box size and vBCs.  Second, the pressure bumps associated these zonal flows are too weak to trap particles, and thus are not directly related to the particle clumping mechanism of the SI.

\subsubsection{Particle Ring Spacing}
\label{subsubsec:spacing}
The left panels of Figures \ref{fig:foSigma}, \ref{fig:peSigma}, and \ref{fig:reSigma} show space-time plots of the azimuthally-averaged particle surface density,  $\langle\Sigma_{\rm p}\rangle_{y}(x)$, relative to $\Sigma_{{\rm g},0}$, the initial gas surface density.  These three figures present outflow, periodic and reflecting vBCs, respectively.
The evolution is similar in all cases.  Radial structures begin to develop at $t \sim 3 P$, first in many closely spaced filaments.  Over time these filaments merge, and the simulations end with a small number of dense filaments, with only a single filament in some of the radially narrow boxes.  The spacing of filaments clearly depends on time and on box size, as addressed below.

First we address the radial drift.  The oscillations in the radial location of the filaments occur on a timescale of one orbital period, which reflect epcicyclic oscillations.  Drift speeds vary between filaments, with denser filaments drifting slower.  When a filament is disrupted, partially or totally, the escaping particles drift at higher speeds, approaching the value of a test particle. In the NSH equilibrium state, the radial drift speed of particles is
\begin{equation}
  v_x = -\frac{2\uptau_{\rm s}}{(1+\mu^2)^2+\uptau_{\rm s}^2}\eta v_{\rm K}, \label{eq:NSHvx}
\end{equation}
where $\mu \equiv \rho_{\rm p}/\rho_{\rm g}$. For the case of a test particle ($\mu = 0$),
\begin{equation}
  v_{x,\mathrm{test}} = -\frac{2}{1+\uptau_{\rm s}^2} \eta v_{\rm K} \uptau_{\rm s}.
\end{equation}
The black arrows in those three figures illustrate $v_{x,\mathrm{test}}$, which are almost parallel with the trajectories of escaping particles.  \autoref{eq:NSHvx} also provides a comparison between the radial drift speed of a filament and the equilibrium radial drift speed (represented by a purple arrow) with a certain $\mu$ value (labeled by purple text).  Since there are so many sub-structures inside a filament, the averaged dust-to-gas density ratio, $\mu$, is only of order $\dex{1}$, which corresponds to a width of $\ell \approx 2\xdex{-2}H$.  The oscillations in the radial location of the filaments occur on a timescale of one orbital period, which reflect epcicyclic oscillations.

At late times in our simulations, the spacing between (and the number of) dense particle filaments varies with vBC and box size, as seen by comparing Figures \ref{fig:foSigma}, \ref{fig:peSigma}, and \ref{fig:reSigma}.  It is dynamically interesting that vBCs can affect the growth of large scale particle structures.  However, the late time behavior of our simulations is less relevant for planetesimal formation, as argued above (because if self-gravity were included then fragmentation would occur earlier).  Thus we analyze the clump spacing during the early strong clumping phase of $t = 15\mbox{--}20P$.

To show the radial spacing of particle filament, \autoref{fig:fftpower} plots the Fourier power spectrum of $\langle\Sigma_{\rm p}\rangle_{y}$.  For boxes with larger $L_X$, the power spectrum extends to lower radial wavenumbers, $k_X$.  In the narrowest boxes with $L_X = 0.2H$, the power is largest at the smallest $k_X$, consistent with the existence of a single dominant clump in these simulations  (see Figures \ref{fig:foSigma}, \ref{fig:peSigma}, \ref{fig:reSigma}).  Thus $L_X = 0.2H$ boxes are too narrow to determine the physical spacing between particle filaments.

For wider boxes with $L_X = 0.4H$ and $0.8H$, the power has a maximum at intermediate wavenumbers of $k_X H/(2 \pi) = 7\pm 2$ for $L_X = 0.4 H$ cases and  $k_X H/(2 \pi) = 8 \pm 2$ for $L_X = 0.8 H$ cases.  The corresponding wavelength, i.e. radial distance betweeen peaks, is $\sim 3\eta r$, somewhat larger than the lengthscale, $\eta r$ that is set by pressure gradients.

We thus conclude that boxes with $L_X \geq 6 \eta r$ should adequately capture the formation of multiple (i.e.\ at least 2) particle filaments for any vBC.  For our adopted value of $\eta r/H = 0.05$, this width corresponds to $L_X \gtrsim 0.3H$.  We emphasize that the spacing of filaments will also depend on $\uptau_{\rm s}$, i.e. particle sizes, which we do not explore here.

\subsubsection{SI-induced Zonal Flows}
\label{subsubsec:zonal}

The right columns of Figures \ref{fig:foSigma}, \ref{fig:peSigma}, and \ref{fig:reSigma} plot perturbations to the azimuthally-averaged gas surface density, $\Delta\langle\Sigma_{\rm g}\rangle_{y}$, for the three vBCs.  The amplitude of the perturbations varies with box size and with vertical BC.  The strongest gas density fluctuations are for the outflow conditions in the widest box, Run ou88.  For this run, the amplitude of gas overdensities reaches $10^{-3}$ at late times (\autoref{fig:foSigma}). The periodic and reflecting cases (in Figures \ref{fig:peSigma} \& \ref{fig:reSigma}) only reach an amplitude of $1.5 \times 10^{-4}$.  Though we have previously argued that the late time behavior of our simulations is not the most relevant, here we consider the late behavior for a couple reasons.  First, the ability of the streaming instability to generate these gas perturbations and (as described below) the corresponding zonal flows is not well known.  Second, we want to emphasize that the largest amplitude  gas bumps in our simulations are still not responsible for (and indeed not capable of) trapping particles by reversing the global pressure gradient.

The gas fluctuations typically contain only one radial wavelength per box, indicative of an inverse cascade to the largest scales.  The inverse cascade is strongest and most persistent in the wide box with outflow vBCs (ou88).  The wide periodic and reflecting cases (Run pe88 and re88) also show an inverse cascade, but with more small-scale features and time-varying structures.  The inverse cascade seen for the gas density is not shared by all aspects of the gas flow or particle dynamics.  For instance,  particle filaments and the gas vertical momentum contain significant small scale structure (including axi-symmetric structure), as seen in Figures \ref{fig:snapshot}, \ref{fig:perebox}, and \ref{fig:radial_wave}.

In a zonal flow, large-scale, radial fluctuations in gas surface density give rise to pressure gradients that balance the Coriolis force of a radially-varying azimuthal flow, i.e.\ 
\begin{equation}
  2\rho_{\rm g} \Omega_0 \delta u_y \approx \fracpartial{P}{x} = c_{\rm s}^2 \fracpartial{\rho_{\rm g}}{x}, \label{eq:geobal}
\end{equation}
where $\delta u_y = u_y + 3/2 \Omega_0 x$ is the deviation of the azimuthal flow from Keplerian.  Geostrophically balanced zonal flows arise in simulations of the magnetorotational instability (MRI, \citet{Johansen2009, Bai2014}).  
They tend to grow to very large radial scales, with a maximum scale of  $\sim 6 H$  seen in large box simulations with $L_X$ up to 20 $H$ \citep{Simon2012, Dittrich2013}.  
We do not measure the maximum radial extent of SI-induced zonal flows as this would require wider simulation domains.  

Now we show the large scale gas density perturbations induced by the SI are in fact zonal flows.  Consider axisymmetric gas density fluctuations with a radial wavelength equal to the box width, with i.e. $2 \pi/k_0 = L_X$, as
\begin{equation}
 \delta \rho_{\rm g} =  \langle\rho_{\rm g}\rangle_y - \langle\rho_{\rm g}\rangle_{x,y}  = \Re (\hat{\rho}'_{\rm g}  e^{ik_0 x}),
\end{equation}
where $\langle \cdot \rangle_{y}$ means an azimuthal average,  $\langle \cdot \rangle_{x,y}$ denotes a horizontal average, and $\hat{\rho}'_{\rm g}$ is a Fourier amplitude.  Then \autoref{eq:geobal} gives a zonal flow speed
\begin{equation}
  \delta u_y = \frac{k_0 c_{\rm s}^2}{2 \Omega_0 \langle\rho_{\rm g}\rangle_{xy}} \Re( i  \hat{\rho}'_{\rm g} e^{ik_0 x}) \label{eq:uyi}.
\end{equation}
\autoref{fig:zonal} shows that gas density and azimuthal flow perturbations are indeed in geostrophic balance.

\autoref{fig:zonal} also shows that the zonal flow is too weak to trap particles.  The zonal flow would have to exceed the headwind speed of  $\eta v_{\rm K} = 0.05 c_{\rm s}$ to halt the inward drift of test particles. The zonal flow amplitude is only $0.01 c_{\rm s}$ in \autoref{fig:zonal}, which shows the strongest zonal flow in all our simulations. Thus none of our simulations are close to creating a pressure trap for particles.

To further emphasize that the zonal flows and their associated pressure bumps appear unconnected with particle clustering,  note that (i) the particle clustering is essentially saturated when zonal flows are still growing,  (ii)  the radial spacing of particle filaments is initially much smaller than zonal flow widths and (iii) zonal flows show a dramatic variation with box size and vBC, quite unlike particle clustering.

The fact that particle clumping appears unconnected from pressure traps is important for a better understanding the mechanism of the SI.  Some attempts to intuitively explain the SI  appeal to the creation of zonal flows and pressure traps \citep[e.g.][]{Jacquet2011, Johansen2014}.  Our analysis shows that the full explanation is more complicated, and that the dynamical gas motions are an essential part of the SI trapping mechanism.

Other works have started to explain this complexity. For instance, the (very closely related to the SI) secular drag layer instability of \citet{Goodman2000} is explained by a toy model of overstable oscillations (see also \citealp{Chiang2010}).  \citet{Lin2017} show that many linear drag instabilities in disks (including the SI) are powered by "$PdV$" work caused by out of phase gas pressure and particle density perturbations.    Thus while pressure perturbations are essential to the SI (as they are for all hydrodynamics), any connection between quasi-static particle trapping in pressure bumps and the SI is not obvious.   

\subsection{Hydrodynamics Away from the Midplane}
\label{subsec:flow}

Even though particle-gas interactions are confined to a thin midplane layer, these interactions trigger hydrodynamic activity that extends through the computational domain  (see \autoref{fig:snapshot} and \autoref{fig:perebox}).  The vBCs have a direct and very strong affect on gas flow through the vertical boundaries.  To quantify this behavior, \autoref{tab:paras} lists the time-averaged gas mass flux through vertical boundaries.   Specifically, the table shows the horizontal average of the absolute value of mass flux ($\langle |\rho_{\rm g} u_z| \rangle$) over both the upper and lower boundaries.  Thus for the outflow vBCs, we get the outflow rate (through both boundaries) as $\dot{\Sigma}_{\rm g} = 2\langle |\rho_{\rm g} u_z| \rangle$. For the periodic vBCs, $\langle |\rho_{\rm g} u_z| \rangle$ measures the gas mass flux through the vertical boundary in either direction, from top to the bottom or vice versa.  The direction of this periodic flow fluctuates in time.  For the reflecting vBCs, there is no data since $\langle |\rho_{\rm g} u_z|\rangle = 0$.

As \autoref{tab:paras} shows, the vertical fluxes of the outflow runs are all $\sim \dex{-3}\rho_{{\rm g},0}c_{\rm s}$, and  vary little with box size.  Each timestep,  a small fraction of gas mass ($\sim\dex{-6}$) is lost before being replenished.  Absent this replenishment, the mass loss timescale of $\dex{3} \Omega_0^{-1}$ is not only very short compared to observationally known disk lifetimes, but also physically implausible from the perspective of the energy budget.  In order to overcome the gravitational potential of the star to escape globally, those gas flows need an energy flux of $v_{\rm K}^2 \dot{\Sigma}_{\rm g}$.  However, the energy powering the SI only provides an energy flux of 
\begin{equation}
\left[\frac{4\uptau_{\rm s}}{(1+\epsilon)^2+\uptau_{\rm s}^2}\right] (\eta v_{\rm K})^2 \Sigma_{\rm p} \Omega_0 \approx (\eta v_{\rm K})^2 \Sigma_{\rm p} \Omega_0,
\end{equation}
where $[\cdot]$ is order unity or smaller (see Eq. 22 in \citet{YJ2007}).  The SI is nowhere near sufficient to power outflows on a global scale (by a factor of $\sim 10^{-5}$)\footnote{In fact, even less power is available to drive outflows, since much of the driving energy is dissipated by drag and by the energetics of angular momentum transport (see Eqs. 23 \& 24 in \citet{YJ2007}).}.

These vigorous outflow rates are thus unrealistic.  In our solutions, the outflow is always subsonic, which explains why the rates are not physically realistic.  Outflow rates are only reliable if all critical points (including MHD wave speeds when magnetic fields are included) are inside the computational domain so that the outflow rate is causally disconnected from the vBCs \citep{Fromang2013}.  Unfortunately, an increased vertical extent does not allow us to find a physical outflow solution with a sonic point\footnote{In our experiments, the outflow rates remain subsonic in even taller boxes (at least up to $L_Z=9.6H$).}.  The fundamental limitation appears to be the local geometry of the shearing box.  Local simulations of MRI outflows find that the critical point for fast magnetosonic waves is not inside the computational domain \citep{Bai2013, Bai2013a}.

For the periodic vBCs, the flow of mass through the connected top and bottom boundaries does not attempt to represent physical outflow or inflow.  This artificial connection between the vertical boundaries can introduce numerical artifacts.  As can be seen from \autoref{tab:paras}, the vertical flux in periodic cases are even stronger than that in the outflow solutions. Moreover, these periodic flows generally decrease with increasing $L_Z$, indicating that the vertical boundaries have less effects when placed further away.

While reflecting vBCs have no flow through the vertical boundaries, reflection at these boundaries can still introduce numerical artifacts.  Such effects could be much stronger in shorter boxes, resulting in the most anomalous case (Run re22) in the smallest box in our simulations.

We conclude that all vBCs introduce numerical artifacts which can affect gas motions throughout the simulation domain.  We are fortunate that the clumping of larger ($\uptau_{\rm s} \gtrsim 0.1$) particles in the midplane is not strongly affected by this imperfection.  However, hydrodynamics away from the midplane is important for  understanding  the lofting of small dust.   We thus caution that we cannot present physically robust results for dust lofting   until either (i) different vBCs give convergent behavior or (ii) a physically robust outflow solution -- with all sonic points -- is found.  Resolving this issue is an important topic for future work.

\section{Conclusions}
\label{sec:conclusions}
We study the effect of numerical box sizes and vBCs on the development of non-linear SI.  As part of this test, we implement outflow boundary conditions in ATHENA.  Our main finding is that the particle concentration triggered by the SI is robust to a variety of numerical setups.  This work negelects self-gravity to focus on purely aerodynamical particle clustering caused by the SI.  The subsequent gravitational collapse of dense particle clumps into planetesimals has been studied elsewhere \citep[][etc.]{Johansen2015, Simon2017}. Our detailed conclusions are:
\begin{itemize}
\item[1] Simulations with different vBCs all produce strong particle concentrations with very similar clumping statistics (see Figures \ref{fig:rhop}, \ref{fig:rhop_scale} and \ref{fig:rhop_scale_all}). The only exception is significantly reduced clumping in the smallest box with reflecting vBCs.  The main effect of vBCs on the particle layer is that outflow vBCs produce vertically thinner layers, especially in shorter boxes (see Figures \ref{fig:hp} and \ref{fig:radial_wave}).
\item[2] The choices of vBCs do significantly affect large-scale gas dynamics, especially for the vertical motions (see Figures \ref{fig:snapshot} and \ref{fig:perebox}).  Outflow vBCs lead to strong gas fluxes out of the box.  Similar to previous MRI simulations, these outflow rates are not reliable, due to the inability of local boxes to capture critical points.
\item[3] The Streaming Instability produces radially banded zonal flows (see Figures \ref{fig:foSigma}, \ref{fig:peSigma}, and \ref{fig:reSigma}).  The SI zonal flows are strongest with the outflow vBCs, but always weaker than MRI-induced zonal flows.  SI zonal flows do not produce pressure maxima -- i.e. pressure perturbations do not overcome the background radial pressure gradient -- and thus do not trap particles (see \autoref{fig:zonal}). 
\item[4] Larger simulations boxes lead to enhanced particle concentration on small scales (see \autoref{fig:rhop_scale}).  Since high-density regions can seed gravitational collapse into bound planetesimals, we predict that the mass distribution in self-gravitating simulations should steepen in larger numerical boxes.  
\item[5] The characteristic particle ring spacing is $\sim 3 \eta r = 0.15 H$ (see \autoref{fig:fftpower}),  similar to the findings of \citet{Yang2014} .  To better capture the interactions between neighboring planetesimal-forming filaments, SI simulation boxes should be wide enough to capture at least two filaments (i.e., $6 \eta r = 0.3 H$ for our parameters).
\end{itemize}

Overall, particle clumping by the SI is numerically robust, especially regarding the choice of vBCs.  These results are reassuring because all vBCs are inherently artificial.  Further investigations of vertical gas motions in tall boxes are needed to understand whether the SI can loft small dust towards disk surfaces.

It is particularly important to understand the effect of box size on the planetesimal mass distributions produced in self-gravitating simulations.  Our results suggest that this effect could be significant (point 4, above).
However, \citet{Simon2017} find a fairly universal mass distribution even when different particles sizes produce particle clustering spectra before collapse.    Understanding the numerical robustness of SI simulations is crucial for making comparisons to the size distributions of the asteroids or Kuiper Belt Objects.

\section*{Acknowledgements}

We thank Anders Johansen, Kaitlin Kratter, Chao-Chin Yang, Xuening Bai, Paola Pinilla for useful discussions.  RL acknowledges support from NASA headquarters under the NASA Earth and Space Science Fellowship Program grant 17-PLANET17R-0011. ANY acknowledges partial support from NASA Astrophysics Theory Grant NNX17AK59G and from the NSF through grant AST-1616929.

\added{\software{ATHENA \citep{Stone2008, Bai2010}, 
          Matplotlib \citep{Matplotlib}, 
          Numpy \citep{Numpy},
          PLAN \citep{PLAN}.}}




\bibliographystyle{aasjournal}
\bibliography{refs}



\appendix
\section{Hydrostatic Balance Test}
\label{app:hydrostatic}
\begin{figure}
  \includegraphics[width=0.48 \linewidth]{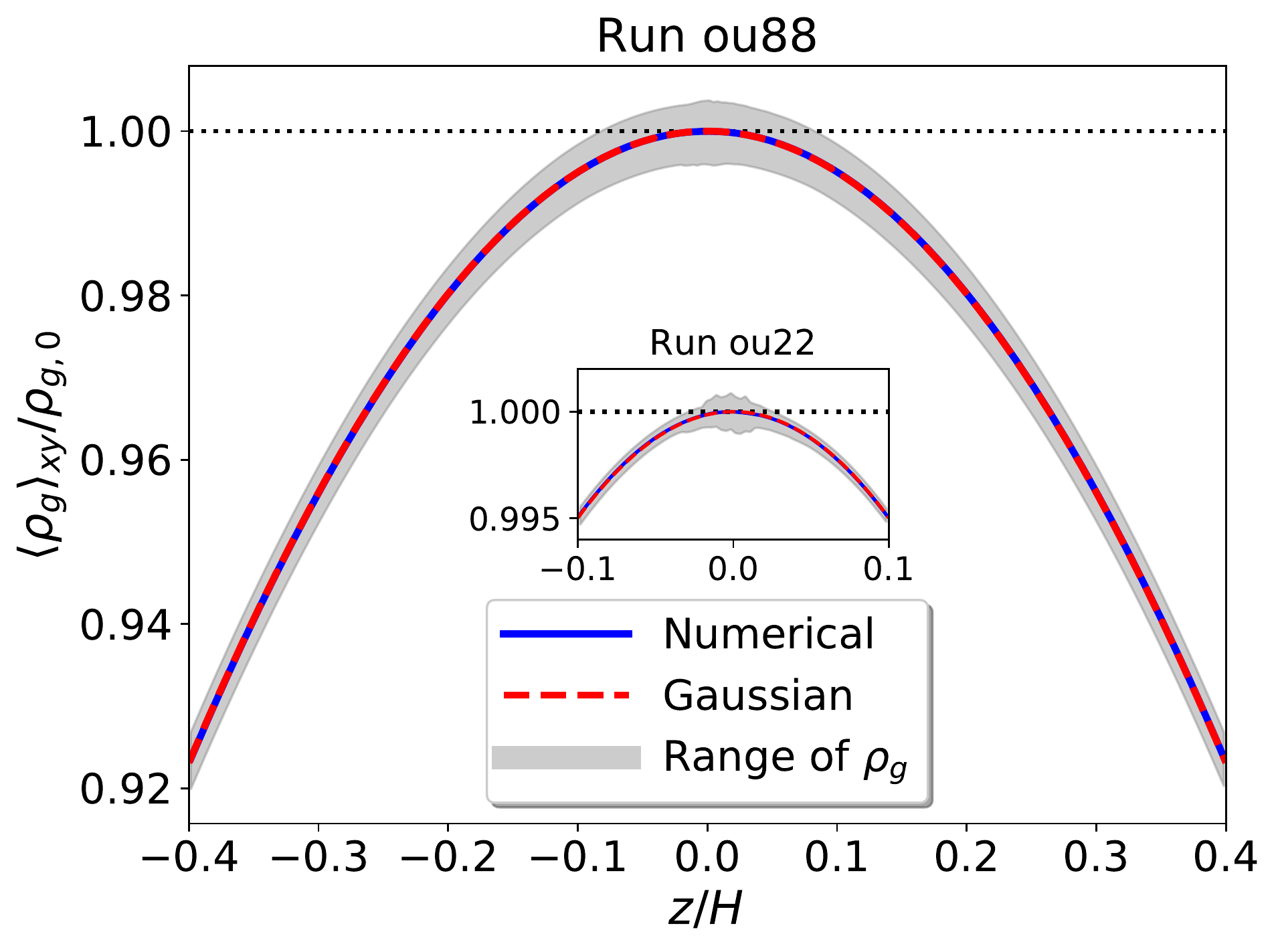}
  \includegraphics[width=0.48\linewidth]{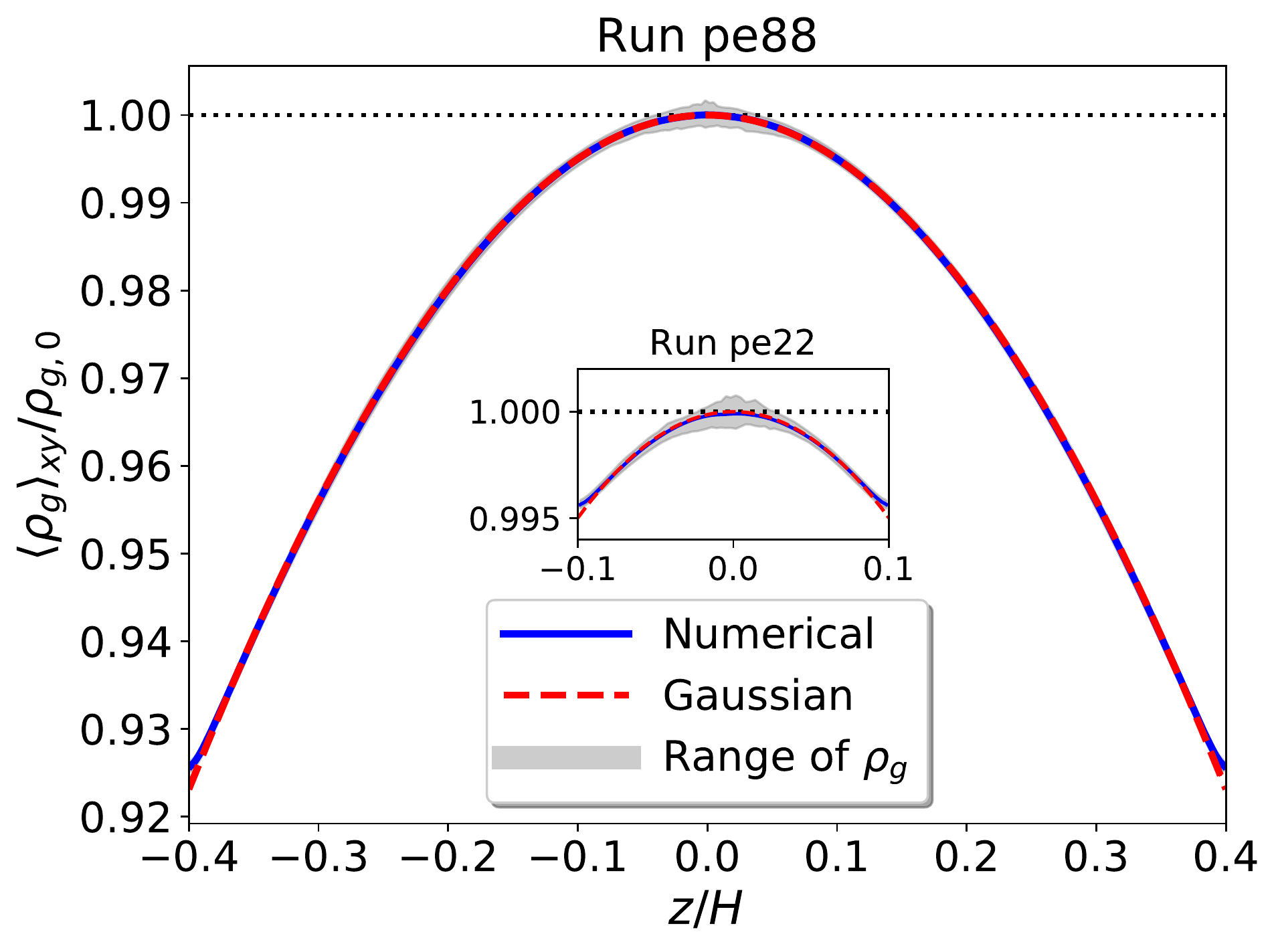} \\
  \includegraphics[width=0.48\linewidth]{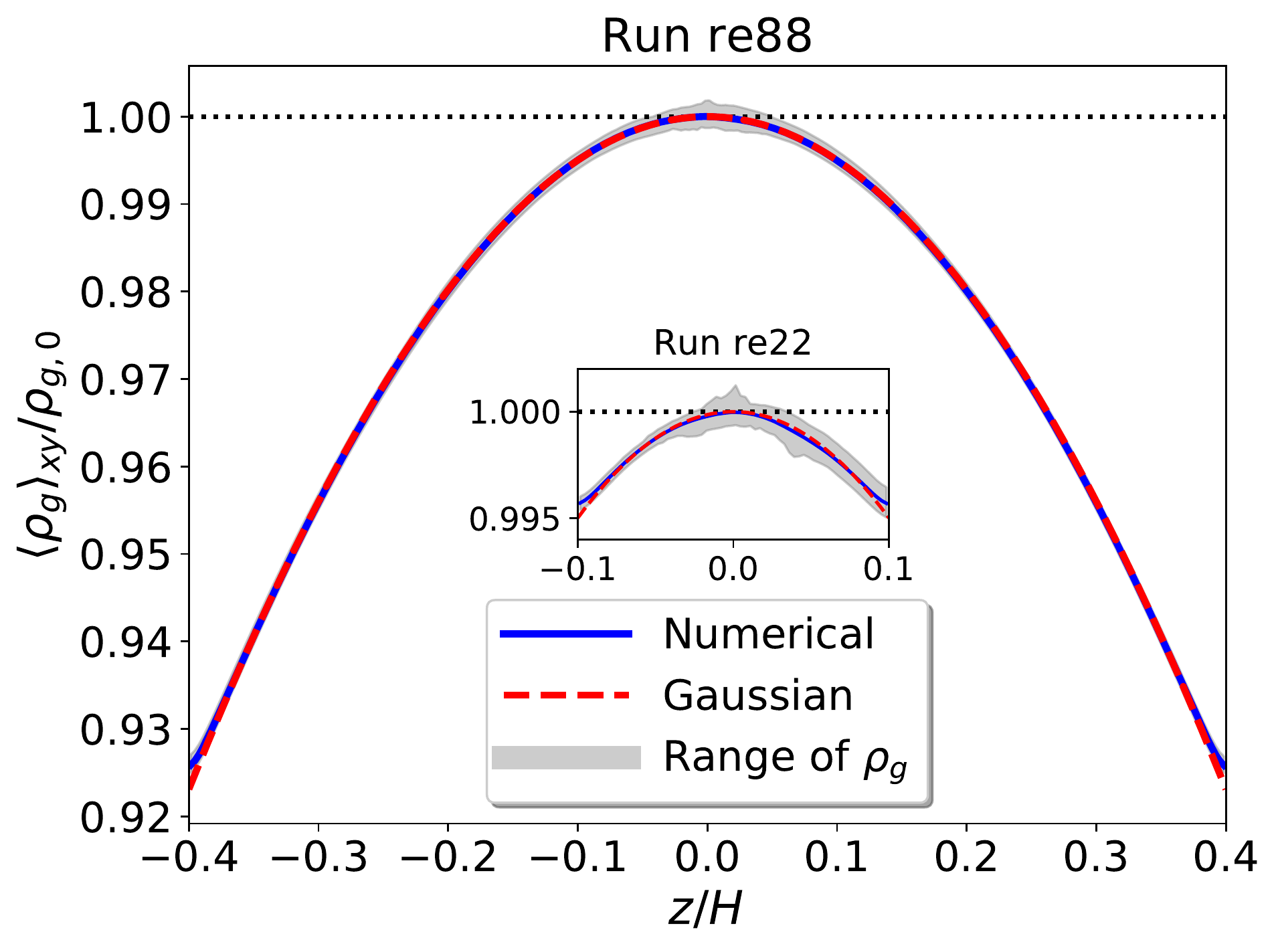}
  \caption{\textbf{\textit{Upper}}: The vertical density structure of gas at the final stage of Run ou88 and the same plot for Run ou22 is embedded (see \autoref{tab:paras}).  The analytical Gaussian profile (\textit{red dash curves}) denotes the initial condition for the gas density in simulations.  \textit{blue solid lines} indicates the horizontally-averaged vertical gas density profile at $t=50P$, which perfectly maintain the initial structure.  The \textit{grey areas} map the range of gas density in the horizontal planes.  The dispersion of gas density reaches the maximum around the midplane as expected due to the interactions between particle layer and gas.  \textbf{\textit{Middle}}: similar to upper panel but for Run pe88 and pe22.  \textbf{\textit{Lower}}: similar to upper panel but for Run re88 and re22.}
 \label{fig:gasVstruc}
\end{figure}

We carefully examine the hydrostatic balance under different vBCs.  \autoref{fig:gasVstruc} shows the comparisons between simulations and analytical result.  In the upper panel, red curves denote the initial analytical Gaussian profile and blue solid lines indicate the horizontally-averaged gas density profile at the final stage of Run ou88/ou22 (the biggest/smallest box under outflow vBCs), which perfectly maintain the initial state after $50$ orbital periods.  The grey areas map the range of gas density in the horizontal planes.  The dispersion of gas density reaches the maximum around the midplane as expected due to the intense interactions between the particle layer and gas.  Such an agreement implies that our implementation of the outflow vBCs is fairly physical.  Similar figures for periodic and reflecting vBCs are presented in the middle and lower panels.  Small deviations exist at the boundaries for periodic and reflecting cases because there is subtle correction applied to vertical gravitational potential in order to use these vBCs.

\end{document}